\documentclass[10pt,journal]{IEEEtran}

\usepackage{graphicx}
\usepackage{orcidlink}
\usepackage{bm}
\usepackage{amsmath,amssymb,amsfonts,mathtools}
\usepackage{booktabs}
\usepackage{array}
\usepackage{multirow}
\usepackage{subfig}
\usepackage{cite}

\DeclareMathOperator*{\argmax}{arg\,max}
\usepackage{url}
\usepackage{microtype}
\usepackage[tableposition=top]{caption}
\usepackage{geometry}
\newgeometry{left=1.4cm,right=1.4cm,bottom=1.2cm,top=1.1cm}

\newcommand{\vect}[1]{\bm{#1}}

\newcommand{\CN}{\mathcal{CN}}

\newcommand{\sinc}{\operatorname{sinc}}

\begin{document}

\title{Micro-Doppler Sensing for UAV Flight-State Estimation in ISAC Networks}

\author{Henglin Pu,  Lu Su, \IEEEmembership{Member, IEEE}, Zhu Han, \IEEEmembership{Fellow, IEEE},
Athina P. Petropulu, \IEEEmembership{Fellow, IEEE}, \\
and Husheng Li, \IEEEmembership{Senior Member, IEEE}
\thanks{H. Pu, L. Su, and H. Li are with the Elmore Family School of Electrical and Computer Engineering, Purdue University, West Lafayette, IN 47907 USA (e-mail: pu36@purdue.edu; lusu@purdue.edu; husheng@purdue.edu).}% \and
\thanks{Zhu Han is with the Department of Electrical Engineering, University of Houston (email: zhan2@uh.edu).} \and
\thanks{Athina P. Petropulu is with the Department of Electrical and Computer Engineering, Rutgers The State University of New Jersey, New Brunswick, NJ, USA
(email: athinap@soe.rutgers.edu). } \and
\thanks{H. Li is also with the School of Aeronautics and Astronautics, Purdue University, West Lafayette, IN 47907 USA (e-mail: husheng@purdue.edu).}%
\and
\thanks{This work was supported by ARO grant W911NF261A271.}
}

\maketitle

\begin{abstract} Integrated sensing and communication (ISAC) provides a promising foundation for low-altitude unmanned aerial vehicle (UAV) monitoring. However, most existing studies focus primarily on position and bulk velocity, thus providing limited information about UAV rotor dynamics and orientation. To address this limitation, this paper develops an orthogonal frequency-division multiplexing (OFDM)-ISAC framework that exploits UAV micro-Doppler signatures to estimate operational states, including rotor rate and rotor-axis orientation, thereby providing richer UAV state information beyond conventional ISAC localization and velocity sensing. The framework progresses from single-station micro-Doppler sensing to multi-station rotor-axis estimation. At each base station, micro-Doppler signatures are first extracted and used to estimate the rotor rate. The framework is then extended to a distributed ISAC network, where view-dependent micro-Doppler signatures from multiple sensing directions are fused to estimate the common rotor-axis direction. Since the fusion relies on Doppler-bandwidth features rather than complex channel phases, inter-station carrier-phase synchronization is not required. We show that three geometrically diverse views can locally identify both the rotor-axis direction and the unknown rotor-tip-speed scale. In addition, a feature-aware radar equation bridges blade radar cross section (RCS), sensing range, transmit power, and observation duration to micro-Doppler detection and rotor-rate estimation accuracy. The effects of residual motion and range migration on the coherent processing interval (CPI) are also analyzed. Simulations and experimental measurements validate the proposed framework and demonstrate the feasibility of rotor-rate estimation and multi-station rotor-axis sensing.
\end{abstract}

\begin{IEEEkeywords}
Integrated sensing and communication (ISAC), micro-Doppler, unmanned aerial vehicle (UAV), rotor-axis estimation, radar cross section (RCS)
\end{IEEEkeywords}

\section{Introduction}
\label{sec:introduction}

Integrated sensing and communication (ISAC) has attracted extensive attention from both industry and academia as a key enabling technology for next-generation wireless networks. By integrating sensing functionality into communication infrastructure, ISAC can transform future wireless networks into pervasive radio sensing platforms while retaining their primary communication functions~\cite{liu2022isac,8999605}. Low-altitude unmanned aerial vehicle (UAV) monitoring is one of the most compelling applications of this paradigm. Existing cellular sites already provide broad coverage and spatial diversity, whereas deploying dedicated radar systems typically requires additional spectrum, hardware, sites, and maintenance~\cite{meng2024uav,mu2023uav}. Therefore, incorporating sensing capabilities into communication infrastructure can complement dedicated radars by providing a dense and wide-area monitoring layer for public safety, airspace management, and low-altitude security.

\begin{table*}[t]
\centering
\captionsetup{font=footnotesize,labelfont=footnotesize}
\caption{Comparison With Representative UAV Micro-Doppler and
Cooperative ISAC Studies}
\label{tab:comparison}

\begin{tabular}{|l|c|c|c|c|c|c|c|c|c|c|c|c|}
\hline
&
\cite{singh2018blade}
&
\cite{ritchie2017multistatic}
&
\cite{gong2022dwell}
&
\cite{costa2024distributed}
&
\cite{costa2025multipropeller}
&
\cite{ma2024frames}
&
\cite{wei2025rotor}
&
\cite{lu2024multibs}
&
\cite{ji2025doppler}
&
\cite{pu2026conference}
&
\cite{Luo2013}
&
\textbf{Proposed}
\\
\hline

Communication-waveform/ISAC sensing
&
&
&
&
$\checkmark$
&
$\checkmark$
&
$\checkmark$
&
$\checkmark$
&
$\checkmark$
&
$\checkmark$
&
$\checkmark$
&
&
$\checkmark$
\\
\hline

Measured-data or hardware validation
&
$\checkmark$
&
$\checkmark$
&
$\checkmark$
&
$\checkmark$
&
$\checkmark$
&
&
$\checkmark$
&
&
$\checkmark$
&
$\checkmark$
&
&
$\checkmark$
\\
\hline

Rotor-rate estimation
&
$\checkmark$
&
&
&
&
&
&
&
&
&
$\checkmark$
&
$\checkmark$
&
$\checkmark$
\\
\hline

Multi-view or distributed sensing
&
&
$\checkmark$
&
&
$\checkmark$
&
$\checkmark$
&
&
&
$\checkmark$
&
$\checkmark$
&
&
$\checkmark$
&
$\checkmark$
\\
\hline

Rotor-axis estimation
&
&
&
&
&
&
&
&
&
&
&
$\checkmark$
&
$\checkmark$
\\
\hline

Rotor-axis identifiability analysis
&
&
&
&
&
&
&
&
&
&
&
&
$\checkmark$
\\
\hline

Estimation bound or uncertainty analysis
&
&
&
&
&
&
&
&
&
&
$\checkmark$
&
$\checkmark$
&
$\checkmark$
\\
\hline

Blade-RCS--range--power analysis
&
&
&
&
&
&
&
&
&
&
&
&
$\checkmark$
\\
\hline

Observation-time analysis
&
&
&
$\checkmark$
&
&
&
&
&
&
&
&
&
$\checkmark$
\\
\hline

Motion/coherence-limit analysis
&
&
&
&
&
&
&
&
&
&
&
&
$\checkmark$
\\
\hline

\end{tabular}

\end{table*}

Most existing ISAC-based UAV studies primarily focus on estimating position and bulk velocity. While these quantities characterize where a UAV is and how its center of mass moves, they do not provide sufficiently rich information about the UAV that could be used to for example  to distinguish UAVs from birds,  reveal hovering drones whose bulk Doppler is buried in clutter, determine whether UAV is carrying a payload, which could bear on threat assessment, or drones maneuvering. This motivates the exploitation of rotor-induced micro-motion as a richer sensing feature. Physically, rotating blades introduce periodic path-length modulation and generate characteristic micro-Doppler signatures, whose spectral spacing reflects the rotor rate and whose extent is determined by the projected blade-tip speed. These features therefore provide valuable information about rotor dynamics, and can also support the detection of hovering UAVs. Extending this idea to a networked ISAC system, spatially separated stations observe different projections of the same rotor motion. By fusing these multi-view micro-Doppler measurements, the common rotor-axis direction and blade-tip-speed scale can be further inferred. Collectively, these micro-Doppler-derived features provide a more informative characterization of UAV dynamic states than localization alone and may further facilitate UAV classification and threat assessment.

However, extracting rotor micro-Doppler signatures is considerably more challenging than detecting the UAV body. Rotor blades generally exhibit much smaller and strongly aspect-dependent radar cross sections (RCS), while the UAV body and surrounding static environment produce much stronger returns that can obscure the weak blade-induced components. This challenge becomes more pronounced at longer sensing ranges, as the received monostatic echo power decays with the fourth power of distance. Consequently, a practical multi-view sensing framework must address two fundamental issues jointly: whether the desired rotor states are geometrically observable from the available sensing views, and whether each participating station can reliably extract the weak micro-Doppler features with sufficient accuracy. This motivates us to investigate the feasibility of micro-Doppler-based rotor-state estimation and to characterize the sensing limits imposed by blade RCS, range, transmit power, and observation duration.

\subsection{Review of Existing Works}

Radar-based counter-UAV sensing does not require target cooperation,
but small UAVs remain difficult to characterize because of their low
altitude, low radial velocity, and small RCS
\cite{park2021antidrone,patel2018review}. The increasing integration of
sensing functionality into cellular infrastructure provides a promising
alternative for wide-area UAV surveillance. Recent studies have
investigated cellular networks for UAV detection, imaging, localization,
and tracking, while recent surveys have highlighted low-altitude UAV
sensing as an important application of cellular ISAC
\cite{song2025overview,bai2026survey}. 

% Measurements
% further show that UAV returns depend strongly on carrier frequency,
% polarization, and viewing aspect
% \cite{rahman2015rcs}. 

% In particular, blade returns
% are generally weaker and more aspect-dependent than body returns.
% Therefore, successful UAV detection or localization does not
% necessarily imply that the rotor-induced features required for
% fine-grained state estimation can be reliably extracted.

% \cite{chen2006microdoppler,chen2019microdoppler}

% \cite{molchanov2014classification,fioranelli2015classification}

Micro-Doppler sensing exploits the frequency modulation induced by
periodic target motion
\cite{chen2006microdoppler,chen2019microdoppler}. For multicopter UAVs,
rotating blades generate harmonic combs, time--frequency flashes, and
varying Doppler extents that support detection and classification
\cite{molchanov2014classification,fioranelli2015classification}.
Micro-Doppler measurements have also been used to estimate physical
rotor parameters such as blade length and rotation rate
\cite{singh2018blade,kang2021spectrum}. However, these approaches
primarily consider a single radar view. A single view observes only a
projection of the blade-tip motion and, when the tip-speed scale is
unknown, cannot independently determine the three-dimensional
rotor-axis direction. Observation duration is also critical for weak
rotor-feature extraction: a longer coherent processing interval (CPI)
improves Doppler resolution and accumulates blade-return energy
\cite{zhang2019resolution}, while experiments have shown that the radar
dwell time must be sufficiently long relative to the rotor period for
blade components to become distinguishable
\cite{gong2022dwell}. 

% Existing studies, however, do not jointly relate
% the required observation duration to blade RCS, sensing range,
% transmit power, desired estimation accuracy, and motion-induced
% coherence limitations.

Communication waveforms have also been exploited directly for UAV
sensing. LTE-based passive-radar systems have demonstrated UAV
detection using both single- and multi-receiver configurations
\cite{dan2019lte,geng2020lte}. More recently, Blandino \emph{et al.} investigated airborne-object
sensing using 5G NR positioning reference signals
\cite{blandino2025airborne} and subsequently evaluated monostatic
gNB sensing for UAV detection and localization under 3GPP-inspired
scenarios \cite{blandino2026gnb}. Real-world 5G-Advanced
base-station measurements have also been used for UAV tracking
\cite{meng2026needle}. These studies provide important evidence that
cellular infrastructure can support practical UAV sensing, especially for localization and trajectory
tracking.

UAV micro-Doppler sensing has more recently been incorporated into
ISAC waveforms. Costa \emph{et al.} developed an experimentally
validated orthogonal frequency-division multiplexing (OFDM)-compatible model for drone-propeller signatures in
distributed ISAC
\cite{costa2024distributed} and subsequently extended it to
multi-propeller drones with bistatic geometry and body reflectivity
\cite{costa2025multipropeller}. Ma \emph{et al.} evaluated the impact
of different 5G frame structures on micro-Doppler-based UAV
identification \cite{ma2024frames}. Wei \emph{et al.} proposed an ISAC
processing method that suppresses body and clutter components before
extracting weak rotor features
\cite{wei2025rotor}, while learning-assisted processing has exploited
micro-Doppler signatures for UAV detection in multipath environments
\cite{yin2025multipath}. Spatial diversity provides a complementary dimension for UAV sensing~\cite{pu2026stf}.
Multistatic micro-Doppler measurements have been combined to improve
UAV payload classification
\cite{ritchie2017multistatic}, while multiple ISAC base stations have
been used to improve UAV position and velocity estimation
\cite{lu2024multibs}. Cooperative cellular sensing has also been
investigated for UAV surveillance and multistatic target-state
estimation \cite{zhang2024perceptive}. Ji \emph{et al.} used bistatic Doppler measurements from
multiple cellular sensing receivers for multistatic UAV trajectory
tracking \cite{ji2025doppler}. These studies demonstrate the feasibility of ISAC-enabled UAV sensing,
including rotor micro-Doppler extraction, and the benefit of geometric
and viewing diversity.  Multi-view micro-Doppler sensing has also been investigated in conventional MIMO and distributed radar systems to recover three-dimensional rotational and precessional motion parameters \cite{Zhao2016MultiAspect,Luo2013,Xu2017}. These methods typically rely on extracting and associating individual micro-Doppler trajectories under structured scattering-center models. For example, \cite{Zhao2016MultiAspect} exploits the relationships among three equivalent scattering centers of a precessing cone, while \cite{Luo2013} reconstructs rotating-scatterer trajectories and discusses the difficulties arising from limited range and time-frequency resolution. Such trajectory-based methods are challenging to extend directly to small UAVs with overlapping, distributed blade echoes. 

Despite these advances, existing UAV-oriented cooperative ISAC studies have primarily focused on detection, localization, trajectory tracking, and classification, leaving three-dimensional rotor-axis estimation largely unexplored. Although conventional multi-view radar studies have investigated rotational-motion estimation, their reliance on individual micro-Doppler trajectories presents challenges for small UAVs. Consequently, several open challenges remain for distributed ISAC networks, including how to recover the UAV rotor-axis direction without individual scattering-center reconstruction, how to characterize the achievable estimation accuracy, and whether weak blade-induced features can be reliably extracted under practical sensing constraints. Building on our conference work on single-station rotor-rate estimation and experimental validation \cite{pu2026conference}, the present work addresses these challenges by exploiting view-dependent Doppler bandwidths for multi-station rotor-axis estimation and deriving the corresponding local identifiability conditions and estimation bounds. Furthermore, while classical radar equations relate received power to range and target RCS \cite{skolnik2008radar}, our analysis connects blade RCS, sensing range, transmit power, and observation duration to rotor-feature estimation requirements, accounting for minimum blade-pass cycles, residual-phase coherence, and range migration. Table~\ref{tab:comparison} summarizes the distinctions from the most closely related studies.

% \cite{skolnik2008radar,richards2014radar}

\subsection{Contributions of the Present Work}

\begin{figure}[!t]
 \centering
 \includegraphics[width=3.0in]{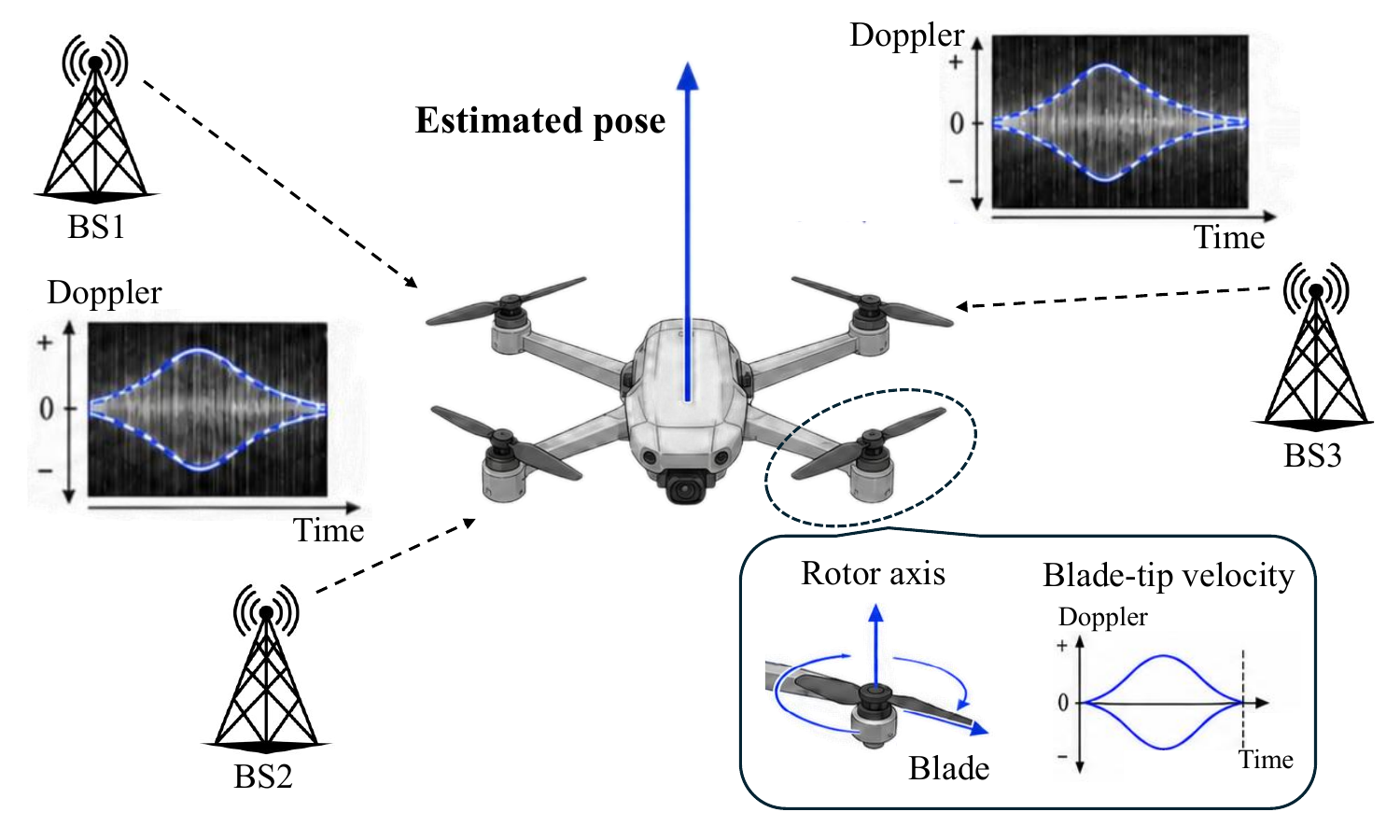}\\
 \caption{Multi-station micro-Doppler rotor-axis sensing. Each station observes a view-dependent blade-tip Doppler span; the extracted signatures are fused to estimate the common rotor-axis direction and tip-speed scale.}
 \label{fig:illustration}
\end{figure}

Motivated by the above research gaps, this paper develops a multi-view OFDM-ISAC framework for UAV rotor-state estimation using blade-induced micro-Doppler signatures, as illustrated in Fig.~\ref{fig:illustration}. The framework combines local rotor-feature extraction with multi-station fusion to estimate rotor dynamics and attitude-related states without requiring inter-station carrier-phase synchronization. In addition to the estimation framework, we investigate the geometric observability and practical sensing feasibility of rotor-state estimation under realistic system constraints. The main contributions are summarized as follows.

\begin{itemize}
\item We develop a unified OFDM-ISAC signal model that incorporates UAV-body translation, individual rotor geometry, blade RCS, body attitude, and blade rotation. The model characterizes both the harmonic structure used for rotor-rate estimation and the time-varying micro-Doppler span used for orientation sensing.

\item We formulate a single-station rotor-rate estimator and extend it to multi-station rotor-axis estimation, enabling joint recovery of rotor dynamics and orientation-related states. With three geometrically diverse views, the estimator recovers two local rotor-axis coordinates together with an unknown Doppler scale. We further derive the corresponding local Fisher information and identify ill-conditioned station geometries.

\item We characterize the observation time required for reliable micro-Doppler detection and rotor-rate estimation as functions of blade RCS, sensing distance, transmit power, and hardware parameters.

\item We build an ISAC hardware prototype and validate the proposed sensing pipeline under both single-station and multi-station configurations. Experimental results demonstrate the practical feasibility of rotor-rate estimation and networked rotor-axis sensing.

\end{itemize}

\subsection{Organization of the Paper}

The remainder of this paper is organized as follows. Section~\ref{sec:model} presents the system model and problem formulation. The single-station algorithm is described in Section~\ref{sec:single}, followed by multi-station rotor-axis estimation in Section~\ref{sec:multi}. Section~\ref{sec:power} derives the RCS--range--power--observation-time relationship. Numerical and experimental results are reported in Section~\ref{sec:results}, and finally Section~\ref{sec:conclusion} concludes the paper.

\section{System Model}
\label{sec:model}

We consider a distributed ISAC network comprising $M_{\rm s}$ monostatic sensing stations, indexed by
$\mathcal M=\{1,\ldots,M_{\rm s}\}$. Station $m$ is located at the known world-frame position $\vect r_m\in\mathbb{R}^3$ and uses an OFDM waveform for communication and sensing. 

Let $\vect p(t)\in\mathbb{R}^3$ denote the UAV center position. The range and line-of-sight (LoS) unit vector from station $m$ to the UAV center are
\begin{equation}
R_m(t)=\|\vect p(t)-\vect r_m\|_2,
\qquad
\vect\ell_m(t)=
\frac{\vect p(t)-\vect r_m}{R_m(t)}.
\label{eq:los}
\end{equation}
Each station locally extracts rotor-related features from its received echo. The time-stamped features, their estimated uncertainties, and the corresponding LoS vectors are subsequently transmitted to a fusion center for rotor-axis estimation. We assume that the distributed ISAC network employs a coordinated sensing scheduler, with sensing resources separated in frequency or code to mitigate inter-station interference, and that each station has sufficient regularly spaced OFDM observations for local micro-Doppler extraction. Although the proposed fusion does not require inter-station carrier-phase synchronization, the measurement-time offsets are assumed to be sufficiently small relative to the UAV rotor-axis variation timescale so that the rotor-axis direction and tip-speed scale remain approximately constant across station measurements.

\subsection{Transmitted OFDM Signal Model}

Each station employs an OFDM waveform with $N$ subcarriers. For even $N$, define the centered subcarrier-index set $\mathcal N=
\left\{-\frac{N}{2},\ldots,\frac{N}{2}-1\right\}$.
The subcarrier spacing is $\Delta f$, the useful OFDM-symbol duration is $T_u=1/\Delta f$, and the cyclic-prefix duration is $T_{\rm cp}$. The total OFDM-symbol interval is $T_s=T_u+T_{\rm cp}.$ Let $X_m[n,k]$ denote the symbol transmitted by station $m$ on subcarrier $n\in\mathcal N$ during OFDM symbol $k$. The symbols are normalized such that
\begin{equation}
\mathbb{E}\{|X_m[n,k]|^2\}=P_{t,m},
\end{equation}
where $P_{t,m}$ is the average conducted transmit power. The complex baseband waveform during the useful part of symbol $k$ is
\begin{equation}
x_{m,k}(\tau)
=
\frac{1}{\sqrt N}
\sum_{n\in\mathcal N}
X_m[n,k]e^{j2\pi n\Delta f\tau},
\qquad 0\leq\tau<T_u.
\label{eq:ofdm_tx}
\end{equation}
The useful bandwidth is approximately $B=N\Delta f$, giving the monostatic range resolution
\begin{equation}
\Delta R\approx\frac{c}{2B}.
\label{eq:range_resolution}
\end{equation}
For a coherent block of $K$ OFDM symbols, the coherent processing interval is $T_{\rm CPI}=KT_s$, the nominal Doppler resolution is $1/T_{\rm CPI}$, and the unambiguous slow-time Doppler interval is
$[-1/(2T_s),1/(2T_s))$.

\subsection{UAV Kinematics and Reflection Channel}

Let $\vect R(t)\in\mathrm{SO}(3)$ denote the rotation matrix that
transforms coordinates from the UAV body frame to the world frame,
where $\mathrm{SO}(3)$ denotes the special orthogonal group of
$3\times3$ proper rotation matrices. Hence,
$\vect R^{\mathsf T}(t)\vect R(t)=\vect I_3$ and
$\det(\vect R(t))=1$. The UAV contains $Q$ rotors. Rotor $q$ has body-frame hub position $\vect h_q\in\mathbb{R}^3$, $B_q$ uniformly spaced blades, inner and outer blade radii $r_{q,1}$ and $r_{q,2}$, and rotation phase $\phi_q(t)$. The signed angular velocity of rotor $q$ is given by
\begin{equation}
\Omega_q(t)=\dot{\phi}_q(t).
\end{equation}
Under the adopted convention, $\Omega_q(t)>0$ denotes rotation according
to the right-hand rule about the positive body-frame $z$-axis, whereas
$\Omega_q(t)<0$ denotes rotation in the opposite direction.

We define the body-frame unit vectors as
\[
\vect e_x=[1,0,0]^{\mathsf T},\qquad
\vect e_y=[0,1,0]^{\mathsf T},\qquad
\vect e_z=[0,0,1]^{\mathsf T}.
\]
For blade $b\in\{0,\ldots,B_q-1\}$ of rotor $q$, its instantaneous phase is given by
\begin{equation}
\phi_{q,b}(t)=\phi_q(t)+\frac{2\pi b}{B_q}.
\end{equation}
A scattering point at radial position
$\rho\in[r_{q,1},r_{q,2}]$ on this blade has world-frame position as 
\begin{align}
\vect x_{q,b}(t,\rho)
={}&
\vect p(t)+\vect R(t)
\left[
\vect h_q+
\vect R_z\!\left(\phi_{q,b}(t)\right)
\rho\vect e_x
\right],
\label{eq:blade_position}
\end{align}
where $\vect R_z(\cdot)$ denotes rotation about the body-frame $z$-axis.

For a multicopter whose dominant rotors share a common normal, the world-frame rotor-axis direction is
\begin{equation}
\vect a(t)=\vect R(t)\vect e_z,
\qquad
\|\vect a(t)\|_2=1.
\label{eq:axis}
\end{equation}
The instantaneous one-way distance from station $m$ to the blade scattering point is given by
\begin{equation}
R_{m,q,b}(t,\rho)
\triangleq
\|\vect x_{q,b}(t,\rho)-\vect r_m\|_2.
\label{eq:blade_range}
\end{equation}

When the UAV dimensions are small relative to $R_m(t)$, a first-order far-field expansion gives
\begin{align}
R_{m,q,b}(t,\rho)
\simeq{}&
R_m(t)+\vect\ell_m^{\mathsf T}(t)\vect R(t)
\nonumber\\
&\times
\left[
\vect h_q+
\vect R_z\!\left(\phi_{q,b}(t)\right)
\rho\vect e_x
\right].
\label{eq:farfield_range}
\end{align}
Therefore, the channel phase contains information about the UAV position, attitude, rotor locations, and blade rotation. Let $t_k=kT_s$, $R_m[k]=R_m(t_k)$, and
$R_{m,q,b}[k,\rho]=R_{m,q,b}(t_k,\rho)$. The frequency-domain channel observed at station $m$ is modeled as
\begin{align}
H_m[n,k]
={}&H_{m,\mathrm{cl}}[n,k]
+\alpha_{m,0}[k]
e^{-j\frac{4\pi f_n}{c}R_m[k]}
\nonumber\\
&+\sum_{q=1}^{Q}\sum_{b=0}^{B_q-1}
\int_{r_{q,1}}^{r_{q,2}}
g_{m,q,b}(\rho,k)
e^{-j\frac{4\pi f_n}{c}R_{m,q,b}[k,\rho]}
\,d\rho ,
\label{eq:uav_channel}
\end{align}
where $f_n$ is the subcarrier frequency, $H_{m,\mathrm{cl}}[n,k]$ is the environmental-clutter response,
$\alpha_{m,0}[k]$ is the dominant UAV-body coefficient, and
$g_{m,q,b}(\rho,k)$ is the distributed blade-scattering coefficient. The coefficient $g_{m,q,b}(\rho,k)$ incorporates propagation loss, antenna gains, polarization mismatch, shadowing, and aspect-dependent blade RCS. The factor $4\pi f_n/c$ in \eqref{eq:uav_channel} accounts for monostatic round-trip propagation.

After bulk-Doppler compensation, where the bulk Doppler refers to the
Doppler shift caused by the translational radial motion of the UAV center,
we assume that the UAV attitude and LoS vector remain approximately
constant within one local processing window. The velocity of a scattering
point at radial position $\rho$ induced by blade rotation is then
\begin{equation}
\frac{d}{dt}
\left[
R_z(\phi_{q,b}(t))\rho\vect e_x
\right]
=
\Omega_q(t)\rho
R_z(\phi_{q,b}(t))\vect e_y .
\label{eq:blade_velocity}
\end{equation}
Under the phase convention in \eqref{eq:uav_channel}, the corresponding
rotor-induced micro-Doppler frequency is
\begin{equation}
f_{\mu,m,q,b}(t,\rho)
=
-\frac{2\Omega_q(t)\rho}{\lambda}
\vect\ell_m^{\mathsf T}(t)R(t)
R_z(\phi_{q,b}(t))\vect e_y,
\label{eq:instantaneous_md}
\end{equation}
which describes the Doppler contribution
from a scattering point at an arbitrary radial position $\rho$ along the
blade. As a result, the entire blade contributes to the received micro-Doppler
signature. Since the Doppler magnitude increases linearly with $\rho$,
its maximum value is attained at the outer blade radius
$\rho=r_{q,2}$. We therefore define the blade-tip speed of rotor $q$ as
\begin{equation}
v_{q,\mathrm{tip}}(t)
=
|\Omega_q(t)|r_{q,2}.
\end{equation}
The blade-tip speed is used to characterize the maximum Doppler extent
of the distributed blade return, rather than to assume that scattering
occurs only at the blade tip.

Since the rotor-induced Doppler varies between positive and negative
values about the bulk-Doppler center, we define the micro-Doppler
half-bandwidth $d_{m,q}(t)$ as the maximum absolute rotor-induced
Doppler offset. Maximizing the magnitude of
\eqref{eq:instantaneous_md} over the blade phase and radial position
gives
\begin{equation}
d_{m,q}(t)
=
\frac{2v_{q,\mathrm{tip}}(t)}{\lambda}
\sqrt{
1-\left[
\vect\ell_m^{\mathsf T}(t)\vect a(t)
\right]^2
}.
\label{eq:bandwidth_model}
\end{equation}
Consequently, the rotor-induced Doppler spectrum ideally extends approximately
from $-d_{m,q}(t)$ to $+d_{m,q}(t)$, corresponding to a total Doppler
span of $2d_{m,q}(t)$.

Because the LoS vector $\vect\ell_m(t)$ differs across sensing stations,
each station observes a different geometric projection of the same rotor
motion and hence measures a different micro-Doppler half-bandwidth.
These view-dependent measurements provide distinct geometric constraints
on the common rotor-axis direction and are jointly fused in the
multi-station estimator.

\subsection{Received OFDM Signal and Feature Mapping}

We assume that the channel delay spread is shorter than $T_{\rm cp}$
and that the channel remains approximately constant during one OFDM
symbol. After cyclic-prefix removal and an $N$-point DFT, the received
signal at station $m$ is
\begin{equation}
Y_m[n,k]
=
X_m[n,k]H_m[n,k]+W_m[n,k],
\label{eq:ofdm_rx}
\end{equation}
where $W_m[n,k]\sim\CN(0,\sigma_{w,m}^2)$ denotes receiver noise.

Because a monostatic station knows its transmitted symbols, the channel
coefficient on an active resource element can be estimated by
\begin{equation}
\widehat H_m[n,k]
=
\frac{Y_m[n,k]}{X_m[n,k]}
=
\frac{X_m^*[n,k]Y_m[n,k]}
{|X_m[n,k]|^2},
\qquad X_m[n,k]\neq 0.
\label{eq:channel_estimate}
\end{equation}

An inverse DFT across the subcarriers transforms the channel estimate into the delay domain:
\begin{equation}
\widehat h_m[r,k]
=
\frac{1}{N}
\sum_{n\in\mathcal N}
\widehat H_m[n,k]e^{j2\pi nr/N},
\quad r=0,\ldots,N-1.
\label{eq:range_profile}
\end{equation}
The delay associated with bin $r$ is $r/(N\Delta f)$, corresponding to the monostatic range
\begin{equation}
R_r=\frac{cr}{2N\Delta f}.
\end{equation}

Let $\mathcal G_m[k]$ denote a narrow range gate centered on the UAV range supplied by the localization layer. The range-focused slow-time signal is
\begin{equation}
z_m[k]
=
\sum_{r\in\mathcal G_m[k]}
\widehat h_m[r,k].
\label{eq:slowtime}
\end{equation}
From \eqref{eq:uav_channel}, $z_m[k]$ contains the UAV-body return, blade-induced micro-Doppler components, residual in-gate clutter, and receiver noise.

After bulk-Doppler compensation and body/clutter suppression, the resulting slow-time signal, denoted by $\widetilde z_m[k]$, is transformed into a time--frequency representation. Two features are then extracted. First, the periodic spectral structure provides the rotor rate $f_{r,q}=\frac{|\Omega_q|}{2\pi}$
.
For $B_q$ approximately identical and uniformly spaced blades, the dominant spectral spacing is the blade-pass frequency $B_qf_{r,q}$.

Second, let $\widehat f_m^+[i]$ and $\widehat f_m^-[i]$ denote the upper and lower Doppler boundaries of the extracted rotor-induced micro-Doppler spectrum in processing window $i$ after bulk-Doppler centering. The estimated half-bandwidth is calculated as
\begin{equation}
\widehat d_m[i]
=
\frac{\widehat f_m^+[i]-\widehat f_m^-[i]}{2}.
\label{eq:feature_map}
\end{equation}
The detailed rotor-rate and bandwidth extraction procedures are presented in Section~\ref{sec:single}. According to \eqref{eq:bandwidth_model}, $\widehat d_m[i]$ is a view-dependent measurement of the rotor-axis direction and rotor-tip-speed scale. Note that since the fusion relies merely on the bandwidth measurements rather than complex channel phases, carrier-phase coherence among the distributed RF front ends is not required.

\subsection{Estimation Formulation}

Based on the preceding signal and channel models, we consider three related estimation problems.

First, we can represent the body-suppressed slow-time signal at station
$m$, over a local coherent processing interval, as
\begin{equation}
\widetilde z_m[k]
=
c_m[k]
+
\sum_{q=1}^{Q}
\sum_{h\in\mathcal H_q}
\alpha_{m,q,h}
e^{j2\pi h f_{r,q}kT_s}
+
\widetilde w_m[k],
\label{eq:single_station_problem_model}
\end{equation}
where $c_m[k]$ represents residual body and clutter leakage,
$\widetilde w_m[k]$ is the processed receiver noise,
$\alpha_{m,q,h}$ is the complex amplitude of harmonic $h$, and
$\mathcal H_q\subset\mathbb{Z}\setminus\{0\}$ is the set of observable signed harmonic orders. Under the identical-blade model, the dominant elements of $\mathcal H_q$ are integer multiples of $B_q$. The single-station problem is to estimate the rotor rates $\{f_{r,q}\}_{q=1}^{Q}$ and the projected micro-Doppler bandwidths despite these nuisance components.

Second, consider a fusion interval during which the dominant rotors share a common axis $\vect a$ and approximately common tip speed $v_{\rm tip}$. We define the Doppler scale as
$\kappa=\frac{2v_{\rm tip}}{\lambda}$.
Suppressing the processing-window index for clarity, the bandwidth reported by station $m$ follows
\begin{equation}
\widehat d_m
=
\kappa
\sqrt{1-(\vect\ell_m^{\mathsf T}\vect a)^2}
+\epsilon_{d,m},
\qquad
\|\vect a\|_2=1,
\label{eq:multiview_problem_model}
\end{equation}
where $\epsilon_{d,m}$ is the bandwidth-estimation error. The multi-station problem is to estimate $(\vect a,\kappa)$ from
$\{\widehat d_m,\vect\ell_m\}_{m=1}^{M_{\rm s}}$ and determine the sensing geometries under which these parameters are locally identifiable. Because \eqref{eq:multiview_problem_model} depends on the UAV attitude only through $\vect a$, rotation about the common rotor axis is not observable from bandwidth-only measurements.

Third, each participating station must receive sufficient blade-echo energy to extract an accurate bandwidth. Over an interval in which the range and link parameters are approximately constant, the coherently accumulated blade-feature SNR is calculated as
\begin{equation}
\gamma_m(T_o)
=
\frac{
P_{t,m}G_{t,m}G_{r,m}\lambda^2\sigma_{b,m}
}{
(4\pi)^3R_m^4L_mk_{\rm B}T_0F_m
}
T_o\eta_m(T_o),
\label{eq:feature_feasibility_model}
\end{equation}
where $T_o$ is the observation duration, $G_{t,m}$ and $G_{r,m}$ are the linear transmit and receive antenna gains, $\sigma_{b,m}$ is the effective blade RCS, $L_m\geq1$ is the aggregate system loss, $k_{\rm B}$ is Boltzmann's constant, $T_0$ is the reference noise temperature, $F_m\geq1$ is the linear receiver noise factor, and $0\leq\eta_m(T_o)\leq1$ is the retained coherent-gain factor. All powers, gains, losses, and RCS values in \eqref{eq:feature_feasibility_model} are expressed in linear units. The feasibility problem is to determine the transmit power and observation duration required to meet the desired detection and estimation accuracy while satisfying the residual-motion and range-migration constraints.

The present study assumes that the UAV has already been detected and associated across stations; the station positions and LoS vectors are available from the localization layer; and the dominant rotor axes and rotation speed remain approximately constant within each local processing window. 

\section{Single-Station Micro-Doppler Feature Estimation}
\label{sec:single}

The preceding section established the mapping from the received OFDM signal to the range-focused slow-time sequence $z_m[k]$, as well as the relationship between rotor motion and micro-Doppler bandwidth. We now describe the processing performed locally at each station to extract the rotor rate and view-dependent half-bandwidth.

\vspace{-1mm}

\subsection{Body Suppression and Bandwidth Extraction}

In the experimental implementation, the OFDM frame boundary is first located by correlating the received waveform with the known preamble. Channel estimation and range focusing are then performed according to \eqref{eq:channel_estimate}--\eqref{eq:slowtime}. After range alignment and bulk-Doppler compensation, a first-order moving-target-indication (MTI) filter is applied to the range-focused sequence:
\begin{equation}
\widetilde z_m[k]
=
z_m[k]-z_m[k-1].
\label{eq:mti}
\end{equation}
This operation suppresses stationary clutter and the quasi-static UAV-body component, thereby improving the visibility of the weaker blade return. Equivalently, as implemented in the receiver, slow-time differencing may be applied to $\widehat H_m[n,k]$ before the inverse DFT because the two operations are linear. The MTI response has magnitude
$2|\sin(\pi fT_s)|$. Therefore, spectral components close to zero Doppler are also attenuated.

Let $g[p]$, $p=0,\ldots,L-1$, denote an analysis window of length $L$, $D$ the hop size, and $N_{\rm D}$ the Doppler-DFT length. The local time--frequency representation is given by
\begin{equation}
Z_m[\nu,i]
=
\sum_{p=0}^{L-1}
\widetilde z_m[iD+p]g[p]
e^{-j2\pi\nu p/N_{\rm D}},
\
S_m[\nu,i]=|Z_m[\nu,i]|,
\label{eq:stft}
\end{equation}
where $i$ is the processing-window index and $\nu$ is the centered Doppler-bin index. The associated Doppler frequency is
\begin{equation}
f_\nu
=
\frac{\nu-\lfloor N_{\rm D}/2\rfloor}
{N_{\rm D}T_s}.
\label{eq:doppler_grid}
\end{equation}

The half-bandwidth estimation in \eqref{eq:bandwidth_model} follows the supplied implementation. For each time window, the search is restricted to $|f_\nu|\leq f_{\max}$, and a guard region $|f_\nu|<f_{\rm g}$ is excluded when locating the dominant blade ridge. After the initial ridge is detected, subsequent searches are confined to a neighborhood of the preceding ridge location to prevent isolated peaks from producing discontinuous jumps. Then a columnwise median is subtracted from $S_m[\nu,i]$ to estimate and remove the local noise floor. Beginning at the detected ridge, the spectral interval is expanded until it contains a prescribed fraction $\eta_{\rm e}$ of the residual energy. The present implementation uses $\eta_{\rm e}=0.95$. The resulting lower and upper edges are denoted by $\widehat f_m^-[i]$ and $\widehat f_m^+[i]$, respectively, and the half-bandwidth is obtained from \eqref{eq:feature_map}. A short median filter is finally applied across adjacent windows to suppress isolated extraction errors. For visualization and multi-view processing, the measured Doppler half-bandwidth can equivalently be expressed as a projected blade-tip velocity $\widehat v_{\perp,m}[i]
=
\frac{\lambda}{2}\widehat d_m[i]$.

\subsection{Rotor-Rate Estimation}

For rotor $q$ with $B_q$ approximately identical and uniformly spaced blades, the observed scattering pattern repeats approximately $B_q$ times during each revolution. Its blade-pass frequency (BPF) is therefore written as
\begin{equation}
f_{{\rm BPF},q}
=
B_q f_{r,q},
\qquad
f_{r,q}
=
\frac{|\Omega_q|}{2\pi},
\label{eq:bpf_relation}
\end{equation}
where $f_{r,q}$ is the rotor rate in revolutions per second. The BPF does not reveal the sign of $\Omega_q$ because the estimator uses spectral magnitudes. In practice, the periodic envelope arises from aspect-dependent blade flashes and interference among the body and blade returns.

The implemented estimator first forms the power envelope
\begin{equation}
e_m[k]=|\widetilde z_m[k]|^2.
\label{eq:power_envelope}
\end{equation}
A moving average is removed from $e_m[k]$ to suppress slow amplitude variations. Let $\overline e_m[k]$ denote the detrended envelope and $w_{\rm H}[k]$ a Hann window spanning $K_o$ slow-time samples. The envelope periodogram is given by
\begin{equation}
P_m(f)
=
\left|
\sum_{k=0}^{K_o-1}
\overline e_m[k]w_{\rm H}[k]
e^{-j2\pi fkT_s}
\right|^2.
\label{eq:envelope_spectrum}
\end{equation}
After smoothing $P_m(f)$ over a narrow frequency interval, its strongest peak $\widetilde f_{p,m}$ is selected within the admissible BPF range $[f_{\min},f_{\max}]$.

Because a higher-order harmonic can be stronger than the fundamental BPF, the estimator examines the subharmonic candidate set
\begin{equation}
\mathcal C_m
=
\left\{
\frac{\widetilde f_{p,m}}{r}:
r=1,\ldots,R_c
\right\}
\cap[f_{\min},f_{\max}],
\label{eq:bpf_candidates}
\end{equation}
where $R_c$ is the maximum tested subharmonic order. Each candidate
$f\in\mathcal C_m$ is then evaluated using the harmonic-comb score
\begin{equation}
J_m(f)
=
\sum_{\substack{h=1\\hf\leq f_{\max}}}^{H_c}
\overline P_m(hf),
\label{eq:harmonic_score}
\end{equation}
where $J_m(f)$ measures the accumulated spectral energy aligned with
the integer harmonics of the candidate frequency $f$, $\overline P_m(f)$ is the smoothed envelope periodogram and $H_c$ is the maximum number of examined harmonics. A larger $J_m(f)$
therefore indicates stronger consistency between the observed spectrum
and the harmonic structure associated with candidate BPF $f$.
The BPF and rotor-rate estimates are then given by
\begin{equation}
\widehat f_{{\rm BPF},m}
=
\argmax_{f\in\mathcal C_m}J_m(f),
\qquad
\widehat f_{r,m}
=
\frac{\widehat f_{{\rm BPF},m}}{B_q}.
\label{eq:rotor_rate_estimator}
\end{equation}
The corresponding rotation speed is calculated as
$\widehat n_{{\rm RPM},m}
=
60\widehat f_{r,m}.$
When several stations observe the same rotor group, their individual estimates can be combined using a median. If multiple rotor groups have resolvable BPFs, the procedure can be employed separately to their spectral-peak families.

Consider a dominant blade scatterer at radius $\rho$ whose coherent slow-time return is
\begin{equation}
u_m[k]
=
\alpha_m
\exp\!\left[
j\beta_m
\sin\!\left(2\pi f_rkT_s+\phi_m\right)
\right]
+v_m[k],
\label{eq:single_scatterer_phase_model}
\end{equation}
where $\alpha_m$ is its complex coefficient, $\phi_m$ is its initial rotation phase, and $v_m[k]\sim\mathcal{CN}(0,\sigma_{v,m}^2)$ is idealized white noise. From the geometry in Section~\ref{sec:model}, the phase-modulation index is
\begin{equation}
\beta_m
=
\frac{4\pi\rho}{\lambda}
\sqrt{1-\left(\vect\ell_m^{\mathsf T}\vect a\right)^2}.
\label{eq:phase_modulation_index}
\end{equation}
Defining the per-sample SNR as
$\gamma_{s,m}=|\alpha_m|^2/\sigma_{v,m}^2$, the many-cycle local Cramér--Rao bound, after accounting for the unknown complex amplitude and initial phase, is
\begin{equation}
\operatorname{var}(\widehat f_r)
\geq
\frac{3}
{\pi^2\beta_m^2T_s^2L(L^2-1)\gamma_{s,m}},
\label{eq:frequency_crb}
\end{equation}
which describes the locally available information in an ideal coherent blade component; it does not imply that the practical envelope-spectrum estimator attains the bound. Multiple blades, aspect-dependent amplitudes, spectral-peak selection, and coherence loss can produce threshold errors that are not captured by a local CRB. The bound is used in Section~\ref{sec:power} only to translate a desired rotor-rate accuracy into an approximate observation-time requirement.

Therefore, a single station provides the rotor rate and a view-dependent projection of the blade-tip speed. It cannot independently separate the true tip-speed scale from the rotor-axis geometry in \eqref{eq:bandwidth_model}.

\section{Multi-Station Rotor-Axis Estimation}
\label{sec:multi}

A single station measures only the product of the rotor-tip-speed scale
and a view-dependent geometric factor. Consequently, neither the
three-dimensional rotor-axis direction nor the true tip speed can be
recovered from one bandwidth measurement. We therefore fuse the
micro-Doppler half-bandwidths obtained from multiple LoS directions.

\subsection{Joint Rotor-Axis and Doppler-Scale Estimation}

Over one fusion interval, the dominant rotors are assumed to share an
approximately common axis $\vect a$ and tip-speed scale $\kappa$, as
defined in \eqref{eq:multiview_problem_model}. Suppressing the
processing-window index, define
\begin{equation}
s_m(\vect a)
=
\sqrt{1-\left(\vect\ell_m^{\mathsf T}\vect a\right)^2},
\qquad
\|\vect a\|_2=1.
\label{eq:aspect_factor}
\end{equation}
The bandwidth measurement at station $m$ is
\begin{equation}
\widehat d_m
=
\kappa s_m(\vect a)+\epsilon_{d,m},
\label{eq:multiview_measurement}
\end{equation}
where $\epsilon_{d,m}$ is the extraction error introduced by noise,
finite Doppler resolution, blade flashing, and imperfect spectral-edge
estimation.

For local analysis, the errors are modeled as independent zero-mean
Gaussian variables with variances $\sigma_{d,m}^2$. Their covariance is
\begin{equation}
\vect\Sigma_d
=
\operatorname{diag}
\left(
\sigma_{d,1}^2,\ldots,\sigma_{d,M_{\rm s}}^2
\right).
\label{eq:bandwidth_covariance}
\end{equation}
This is a working model used for weighted estimation and local
performance analysis. In the simulations, $\sigma_{d,m}$ is prescribed
to evaluate sensitivity to bandwidth error.

Collecting the station measurements, the joint weighted
least-squares estimate is
\begin{equation}
(\widehat{\vect a},\widehat\kappa)
=
\arg\min_{\substack{\|\vect a\|_2=1\\\kappa\geq0}}
\sum_{m=1}^{M_{\rm s}}
w_m
\left[
\widehat d_m-\kappa s_m(\vect a)
\right]^2,
\label{eq:axis_nls}
\end{equation}
where $w_m=\sigma_{d,m}^{-2}$. Equal weights are used when station-specific variances are unavailable,
as in the current three-station implementation.

For a fixed candidate axis, the minimizing scale is given by
\begin{equation}
\widehat\kappa(\vect a)
=
\left[
\frac{
\displaystyle\sum_{m=1}^{M_{\rm s}}
w_m\widehat d_m s_m(\vect a)
}{
\displaystyle\sum_{m=1}^{M_{\rm s}}
w_m s_m^2(\vect a)
}
\right]_+,
\label{eq:kappa_profile}
\end{equation}
where $[x]_+=\max(x,0)$. Substituting
\eqref{eq:kappa_profile} into \eqref{eq:axis_nls} gives
\begin{equation}
\widehat{\vect a}
=
\arg\min_{\|\vect a\|_2=1}
\sum_{m=1}^{M_{\rm s}}
w_m
\left[
\widehat d_m-\widehat\kappa(\vect a)s_m(\vect a)
\right]^2.
\label{eq:profiled_axis_estimator}
\end{equation}
The implemented estimator evaluates this objective over an
azimuth--elevation grid and profiles $\kappa$ analytically at every
candidate direction.

When the rotor rate and effective blade-tip radius are known,
$\kappa$ can instead be fixed according to
\begin{equation}
\kappa
=
\frac{2v_{\rm tip}}{\lambda}
=
\frac{4\pi r_{\rm tip}f_r}{\lambda}.
\label{eq:known_kappa}
\end{equation}

\subsection{Local Identifiability and Accuracy}

The unit vector $\vect a\in\mathbb S^2$ has two local degrees of
freedom, while the unknown scale $\kappa$ contributes a third. Let
$\vect U\in\mathbb R^{3\times2}$ be an orthonormal basis for the tangent
plane at $\vect a$, satisfying
\begin{equation}
\vect U^{\mathsf T}\vect U=\vect I_2,
\qquad
\vect U^{\mathsf T}\vect a=\vect 0.
\end{equation}
For a small tangent perturbation $\vect\delta\in\mathbb R^2$,
\begin{equation}
\vect a(\vect\delta)
=
\vect a+\vect U\vect\delta
+\mathcal O(\|\vect\delta\|_2^2).
\end{equation}

For $0<s_m(\vect a)<1$, the Jacobian row of
$\kappa s_m(\vect a)$ with respect to
$[\vect\delta^{\mathsf T},\kappa]^{\mathsf T}$ is
\begin{equation}
\vect h_m^{\mathsf T}
=
\left[
-\kappa
\frac{\vect\ell_m^{\mathsf T}\vect a}
{s_m(\vect a)}
\vect\ell_m^{\mathsf T}\vect U
\quad
s_m(\vect a)
\right].
\label{eq:axis_jacobian}
\end{equation}
Defining
$\vect H=[\vect h_1,\ldots,\vect h_{M_{\rm s}}]^{\mathsf T}$,
the local Fisher information matrix is
\begin{equation}
\vect J
=
\vect H^{\mathsf T}
\vect\Sigma_d^{-1}
\vect H.
\label{eq:axis_fim}
\end{equation}

Because $(\vect a,\kappa)$ has three local unknowns, at least three
scalar bandwidth measurements are necessary when $\kappa$ is unknown.
The parameters are locally identifiable to first order when $\operatorname{rank}(\vect H)=3$.
Thus, three stations are necessary but are not sufficient under every
placement. Nearly parallel LoS directions provide redundant
measurements and can make $\vect J$ poorly conditioned.

For a locally unbiased estimator, the local parameter covariance
satisfies
\begin{equation}
\operatorname{cov}
\left(
\begin{bmatrix}
\widehat{\vect\delta}\\
\widehat\kappa-\kappa
\end{bmatrix}
\right)
\succeq
\vect J^{-1}.
\label{eq:axis_crb}
\end{equation}
The corresponding lower bound on the mean-squared angular error is
\begin{equation}
\mathbb E\{\theta_a^2\}
\gtrsim
\operatorname{tr}
\left(
[\vect J^{-1}]_{1:2,1:2}
\right).
\label{eq:axis_angular_crb}
\end{equation}
Because $\vect a$ and $-\vect a$ produce identical bandwidths, the
simulation evaluates the acute axis error as
\begin{equation}
\theta_a
=
\cos^{-1}
\left(
\min\left\{
1,
\left|
\widehat{\vect a}^{\mathsf T}\vect a
\right|
\right\}
\right).
\label{eq:acute_axis_error}
\end{equation}
Equations \eqref{eq:axis_crb}--\eqref{eq:acute_axis_error} correspond
directly to the estimator-RMSE and local-CRB curves reported in
Section~\ref{sec:results}.

\subsection{Axial-Rotation Ambiguity}

The bandwidth measurements determine the common rotor-axis direction,
but not the complete UAV attitude. To see this, consider
$\vect R\in\mathrm{SO}(3)$ and $\vect R'=\vect R\vect R_z(\psi),$ where $\psi$ is arbitrary. Since
$\vect R_z(\psi)\vect e_z=\vect e_z$, we have
\begin{equation}
\vect R'\vect e_z
=
\vect R\vect R_z(\psi)\vect e_z
=
\vect R\vect e_z
=
\vect a.
\end{equation}
Therefore, rotation about the common rotor axis leaves all bandwidth
measurements unchanged, regardless of the number of stations. The
recovered axis provides two attitude degrees of freedom associated with
its tilt, whereas rotation about that axis remains unobservable.
Resolving the missing component would require an additional body-fixed
feature, such as a distinguishable arm or rotor hub, which is beyond the
range resolution of the present waveform. Accordingly, this work
estimates only the rotor-axis direction and Doppler scale.

\section{Micro-Doppler Sensing Feasibility}
\label{sec:power}

The preceding sections establish the signal-processing and geometric
conditions for rotor-rate and rotor-axis estimation. These conditions
are useful only when the weak blade return can be observed with
sufficient SNR. In this section, we therefore relate the blade RCS, sensing range,
transmit power, and observation duration, and then determine whether
the required observation can be completed within a motion-limited
coherent processing interval.

\subsection{Blade RCS and Link Budget}

The blade return is strongly dependent on viewing aspect. To obtain an
analytically tractable illustration of this dependence, we approximate
the blade as a thin rectangular plate with effective length $L_b$ and
width $W_b$. Under a physical-optics approximation, its monostatic RCS can be written as
\begin{align}
\sigma_b(\theta,\varphi)
={}&
\eta_{\rm mat}
\frac{4\pi(L_bW_b)^2}{\lambda^2}
\cos^2\theta
\nonumber\\
&\times
\sinc^2\!\left(
\frac{2L_b}{\lambda}
\sin\theta\cos\varphi
\right)
\sinc^2\!\left(
\frac{2W_b}{\lambda}
\sin\theta\sin\varphi
\right),
\label{eq:blade_rcs}
\end{align}
where $\theta$ is the angle between the LoS and the blade normal,
$\varphi$ is the in-plane aspect angle,
$\sinc(x)=\sin(\pi x)/(\pi x)$, and
$0<\eta_{\rm mat}\leq1$ represents material, polarization, and modeling
losses. Equation~\eqref{eq:blade_rcs} is used only as an illustrative
aspect-dependent model; calibrated blade RCS should be used for
quantitative system design.

For a monostatic station at range $R_m$, the received power from a
component with RCS $\sigma_m$ is
\begin{equation}
P_{r,m}
=
\frac{
P_{t,m}G_{t,m}G_{r,m}\lambda^2
\sigma_m
}{
(4\pi)^3R_m^4L_m
},
\label{eq:radar_received_power}
\end{equation}
where $P_{t,m}$ is the conducted transmit power,
$G_{t,m}$ and $G_{r,m}$ are the linear transmit and receive antenna
gains, and $L_m\geq1$ contains propagation and implementation losses.

Let $F_m$ be the linear receiver noise factor, $B_{n,m}$ the equivalent
noise bandwidth, $T_0$ the reference noise temperature, and
$G_{p,m}$ the effective processing gain. The corresponding
post-processing SNR is
\begin{equation}
\gamma_m
=
\frac{
P_{t,m}G_{t,m}G_{r,m}\lambda^2
\sigma_mG_{p,m}
}{
(4\pi)^3R_m^4L_m
k_{\rm B}T_0F_mB_{n,m}
},
\label{eq:post_processing_snr}
\end{equation}
where $k_{\rm B}$ is Boltzmann's constant. All powers, gains, losses,
and RCS values in \eqref{eq:post_processing_snr} are expressed in
linear units.

For a required post-processing SNR $\gamma_{\rm req,m}$,
\eqref{eq:post_processing_snr} gives
\begin{align}
P_{t,m}^{\rm req}
&=
\frac{
\gamma_{\rm req,m}(4\pi)^3R_m^4L_m
k_{\rm B}T_0F_mB_{n,m}
}{
G_{t,m}G_{r,m}\lambda^2
\sigma_mG_{p,m}
},
\label{eq:required_power}\\
R_m^{\max}
&=
\left[
\frac{
P_{t,m}G_{t,m}G_{r,m}\lambda^2
\sigma_mG_{p,m}
}{
\gamma_{\rm req,m}(4\pi)^3L_m
k_{\rm B}T_0F_mB_{n,m}
}
\right]^{1/4}.
\label{eq:maximum_range}
\end{align}
Thus, we have
\begin{equation}
P_{t,m}^{\rm req}\propto\frac{R_m^4}{\sigma_m},
\qquad
R_m^{\max}\propto
\left(P_{t,m}\sigma_m\right)^{1/4}.
\label{eq:link_scaling}
\end{equation}
Every doubling of range therefore requires approximately 12~dB
additional transmit power, while a 10~dB reduction in RCS requires
10~dB additional power at the same range.

The relevant RCS and SNR threshold depend on the sensing task. Body
detection is governed by the relatively strong body RCS, whereas
micro-Doppler extraction is governed by the substantially weaker
dynamic blade RCS remaining after body and clutter suppression. For
multi-station rotor-axis estimation, every participating station must
provide a usable bandwidth measurement in addition to satisfying the
geometric rank condition.

\begin{figure}[t]
\centering
\subfloat[Simulated micro-Doppler spectrum.]{
    \includegraphics[width=0.25\textwidth]{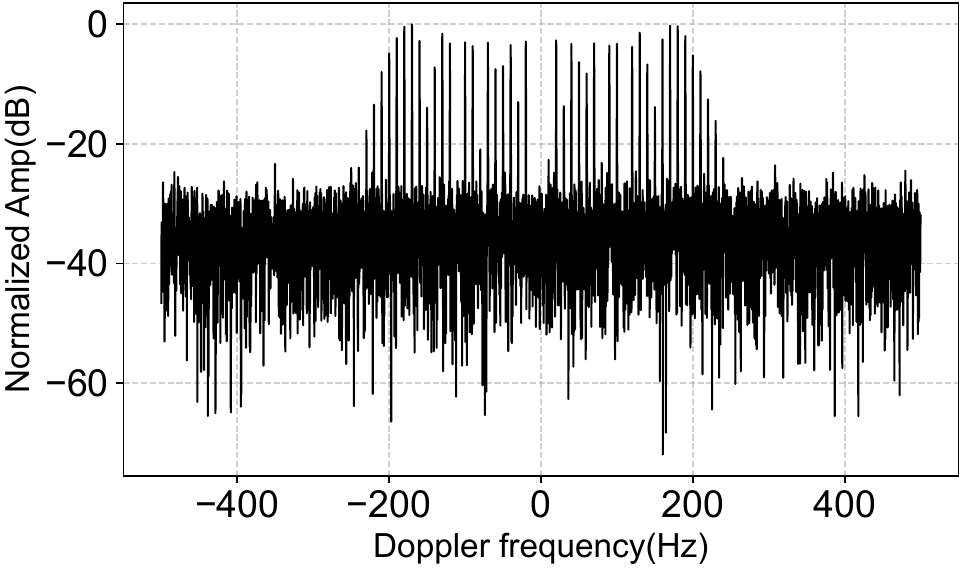}
}\hfill
\subfloat[Frequency-estimation MSE and CRB.]{
    \includegraphics[width=0.21\textwidth]{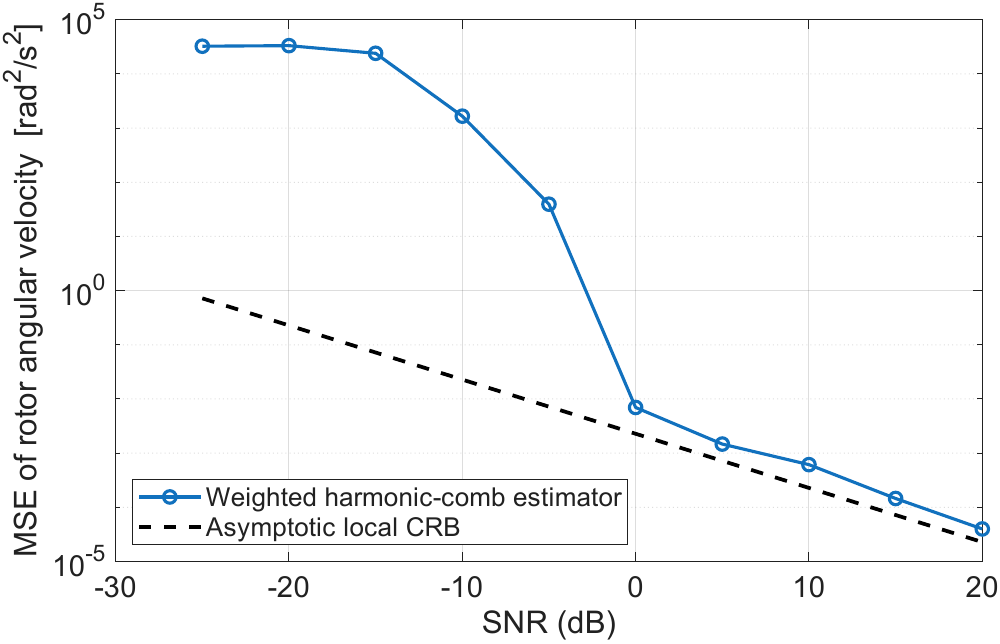}
}
\caption{Single-station rotor-frequency simulation.}
\label{fig:single_sim}
\end{figure}

\subsection{Required Observation Time}

The harmonic structure extracted in Section~\ref{sec:single}
connects the observation-time analysis to classical sinusoid
detection. Although the instantaneous micro-Doppler frequency of
an ideal rotating scatterer varies sinusoidally, its complex
return is phase-modulated and contains multiple harmonics.
Processing matched to a coherent component produces an output
SNR proportional to its received energy and, consequently, to the
observation duration. The required SNR depends on the detector
and the prescribed detection and false-alarm probabilities.
Here, a specified post-processing SNR threshold is used to obtain
an idealized duration benchmark, rather than a calibrated detection
guarantee for the practical envelope-spectrum estimator. We define the blade-echo SNR accumulation rate as
\begin{equation}
\rho_m(R_m,\sigma_{b,m})
=
\frac{P_{r,m}}{N_{0,m}}
=
\frac{
P_{t,m}G_{t,m}G_{r,m}\lambda^2\sigma_{b,m}
}{
(4\pi)^3R_m^4L_m k_{\rm B}T_0F_m
},
\label{eq:snr_accumulation_rate}
\end{equation}
where $N_{0,m}=k_{\rm B}T_0F_m$ is the receiver noise power
spectral density, and $\rho_m$ has units of $\mathrm{s}^{-1}$.
Under ideal coherent processing matched to the blade return
and exploiting its full modeled energy, the accumulated SNR is given by
\begin{equation}
\gamma_{c,m}(T_{\rm o})=\rho_mT_{\rm o},
\label{eq:ideal_coherent_snr}
\end{equation}
where $T_{\rm o}$ is the observation duration. This expression
is equivalent to setting $G_{p,m}=B_{n,m}T_{\rm o}$ in
\eqref{eq:post_processing_snr}. It represents an ideal energy
benchmark, not the gain achieved by directly summing
uncompensated phase-modulated samples. If processing uses only
a subset of the harmonic energy, the corresponding available
power must replace $P_{r,m}$.

\begin{figure}[t]
\centering
\subfloat[Three-station geometry.]{
    \includegraphics[width=0.23\textwidth]
    {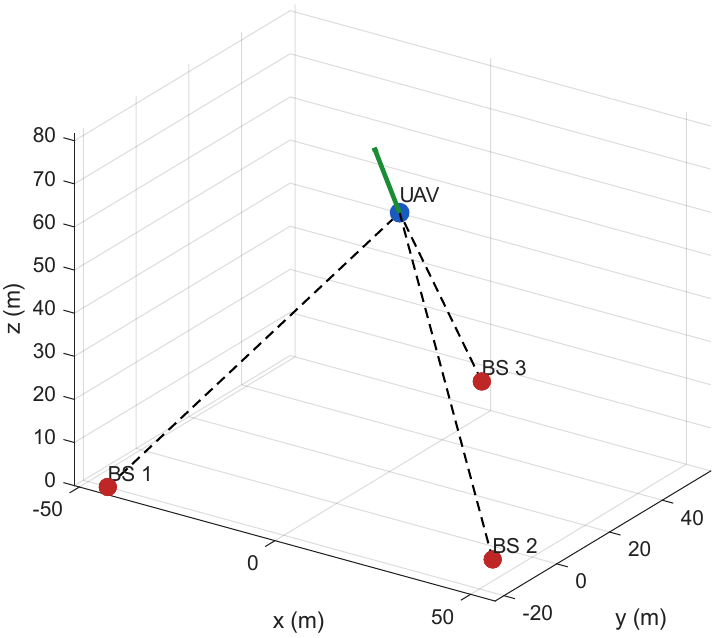}
}\hfill
\subfloat[Joint and known-scale estimation.]{
    \includegraphics[width=0.23\textwidth]
    {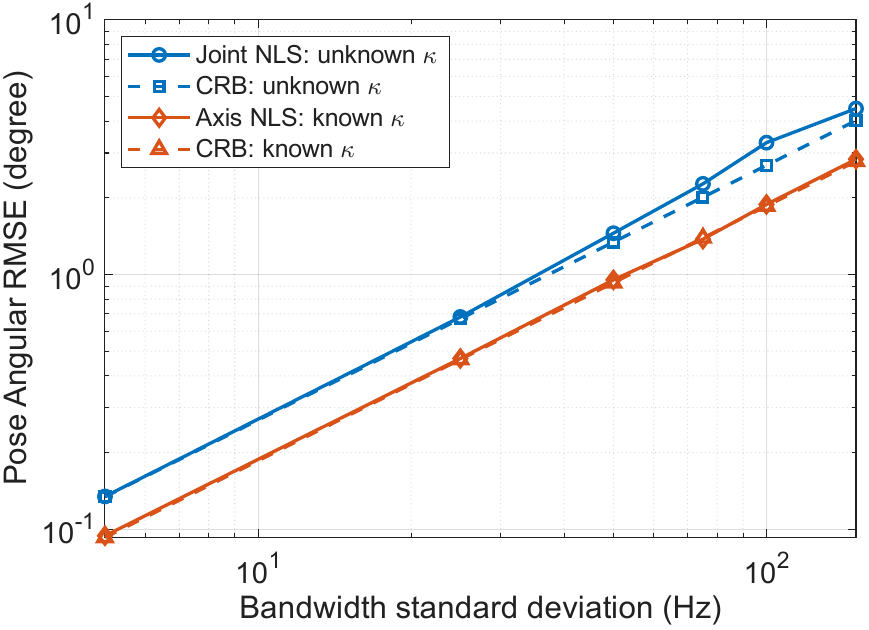}
}
\caption{Multi-station rotor-axis estimation. (a) Station geometry and
true rotor axis. (b) Rotor-axis RMSE and local CRBs when the Doppler
scale is jointly estimated or known.}
\label{fig:multiview}
\end{figure}

The duration needed to reach a specified accumulated
blade-feature SNR $\gamma_{\rm req}$ is therefore
\begin{equation}
\begin{aligned}
T_{{\rm det},m}
&=
\frac{\gamma_{\rm req}}{\rho_m} =
\frac{
\gamma_{\rm req}(4\pi)^3R_m^4L_m k_{\rm B}T_0F_m
}{
P_{t,m}G_{t,m}G_{r,m}\lambda^2\sigma_{b,m}
}.
\end{aligned}
\label{eq:detection_time}
\end{equation}
For fixed threshold, antenna gains, and losses, we have $T_{{\rm det},m}
\propto
\frac{R_m^4}{P_{t,m}\sigma_{b,m}}$,
Thus, doubling the range increases the energy-limited duration
by a factor of sixteen, whereas increasing transmit power or
blade RCS reduces it proportionally. This relationship
characterizes ideal feature detectability; detecting a harmonic
does not by itself guarantee accurate bandwidth extraction.

\begin{figure*}[t]
\centering
\subfloat[Representative time--velocity spectrum.]{
    \includegraphics[width=0.30\textwidth]
    {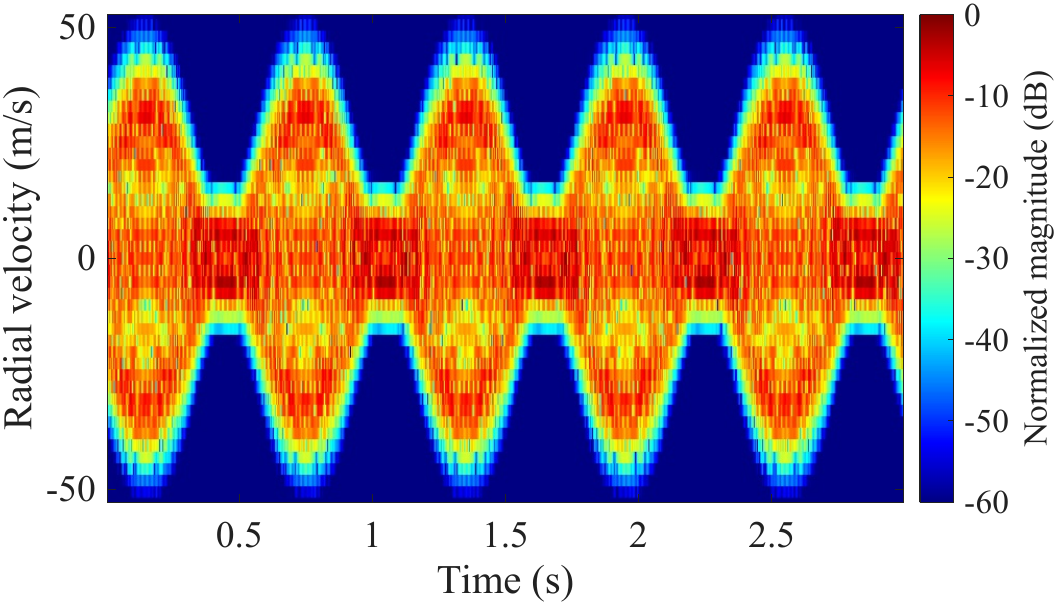}
}
\subfloat[View-dependent half-bandwidths.]{
    \includegraphics[width=0.28\textwidth]
    {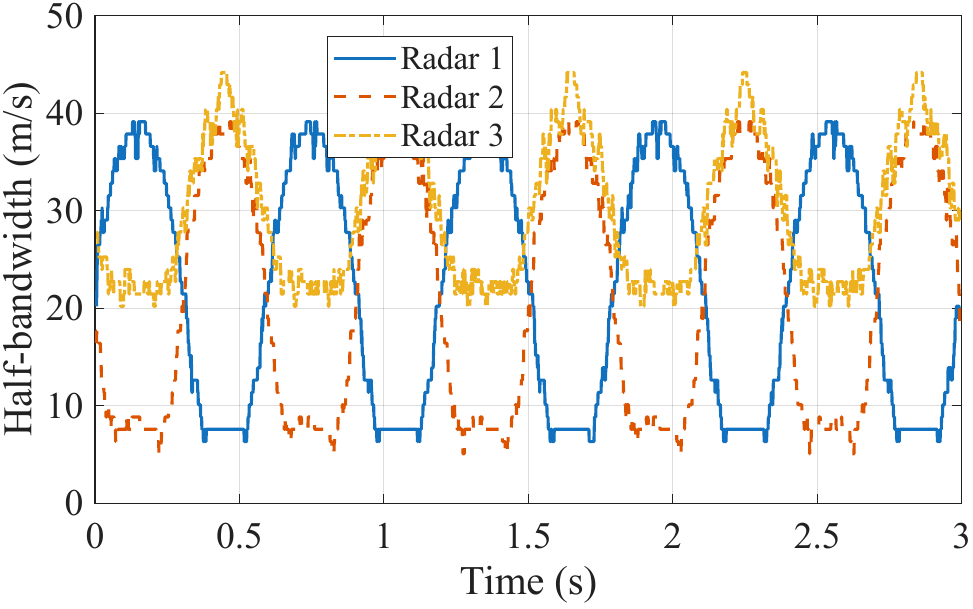}
}
\subfloat[Estimated pitch angle.]{
    \includegraphics[width=0.28\textwidth]
    {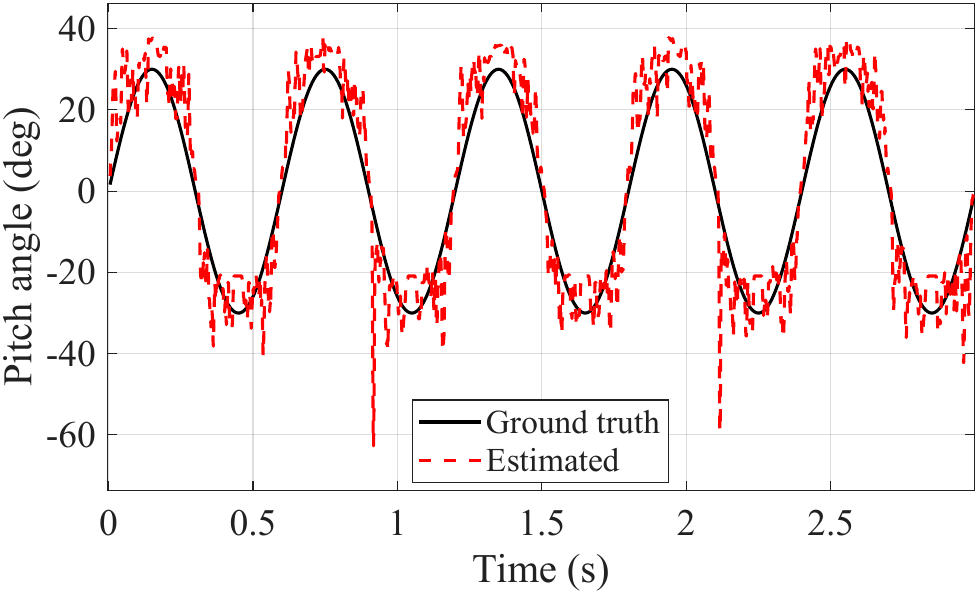}
}
\caption{Signal-level simulation of multi-station rotor-axis estimation:
(a) representative micro-Doppler spectrum, (b) half-bandwidths measured
by the three stations, and (c) recovered pitch angle.}
\label{fig:multiview_sim}
\end{figure*}

\begin{figure}[!t]
\centering
\includegraphics[width=0.72\linewidth]
{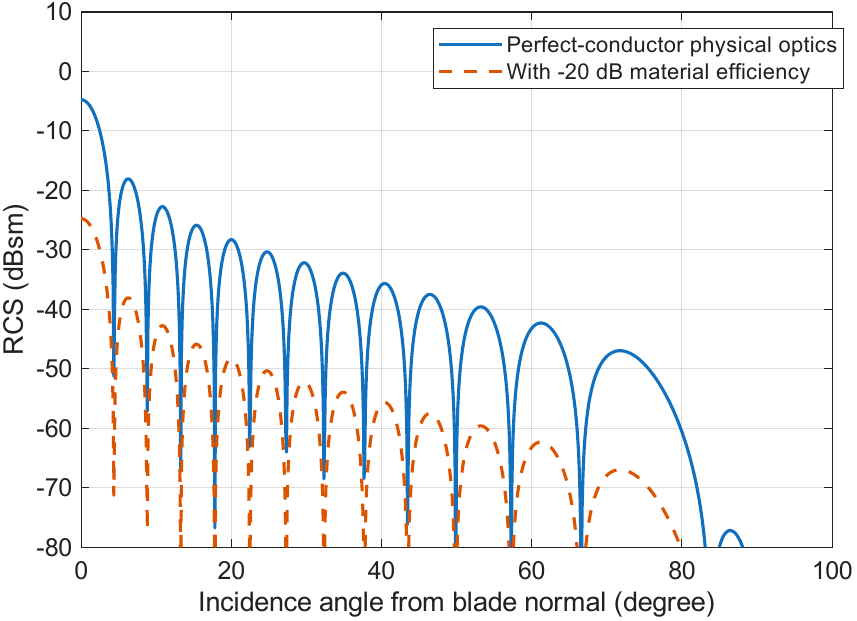}
\caption{Aspect dependence of the rectangular-blade RCS model. The
material-efficiency factor is illustrative and is not a universal
drone-RCS value.}
\label{fig:rcs_aspect}
\end{figure}

A separate requirement follows from rotor-rate estimation
accuracy. For this benchmark, the received power $P_{r,m}$ is
identified with the coherent component power in
\eqref{eq:single_scatterer_phase_model}. Under the equivalent
independent-sample noise normalization, the per-sample SNR is
$\gamma_{s,m}=\rho_mT_s$. Substituting this relation and
$L\simeq T_{\rm o}/T_s$ into \eqref{eq:frequency_crb} gives,
in the many-cycle, large-$L$ regime,
\begin{equation}
\operatorname{var}(\widehat f_r)
\gtrsim
\frac{3}
{\pi^2\beta_m^2\rho_mT_{\rm o}^3},
\label{eq:continuous_time_crb}
\end{equation}
where $\beta_m$ is the phase-modulation index defined in
\eqref{eq:phase_modulation_index}. Requiring this local bound
to be no greater than a target squared rotor-rate RMSE
$\epsilon_f^2$ yields
\begin{equation}
T_{f,m}
=
\left(
\frac{3}
{\pi^2\beta_m^2\rho_m\epsilon_f^2}
\right)^{1/3}.
\label{eq:frequency_observation_time}
\end{equation}
This duration is a local accuracy benchmark, not a sufficient
guarantee that the practical estimator achieves the target RMSE.
For fixed $\beta_m$ and $\epsilon_f$,
\begin{equation}
T_{f,m}
\propto
\frac{R_m^{4/3}}
{(P_{t,m}\sigma_{b,m})^{1/3}}.
\label{eq:frequency_time_scaling}
\end{equation}

A minimum number of blade-pass cycles is also imposed to expose
the periodic rotor signature. For rotor $q$, observing
$N_{\rm cyc}$ blade-pass cycles requires
\begin{equation}
T_{{\rm cyc},q}
=
\frac{N_{\rm cyc}}{B_qf_{r,q}}.
\label{eq:cycle_time}
\end{equation}
Combining these criteria gives the idealized observation-time
design benchmark
\begin{equation}
T_{{\rm req},m}
=
\max\left\{
T_{{\rm det},m},
T_{f,m},
T_{{\rm cyc},q}
\right\}.
\label{eq:total_observation_time}
\end{equation}
Depending on the operating parameters, the duration is limited
by the minimum cycle count, local rotor-rate accuracy, or feature
detection. At sufficiently short range or high power, the cycle
requirement sets a lower floor. At long range or low power,
detection eventually dominates and the required duration follows
the $R_m^4/(P_{t,m}\sigma_{b,m})$ scaling.

\subsection{Motion-Limited Coherent Integration}

The linear accumulation in \eqref{eq:ideal_coherent_snr} assumes that
the predicted blade phase remains coherent throughout the observation.
After compensation using the UAV-body track, let $\Delta v_m$ and
$\Delta a_m$ denote the residual radial-velocity and radial-acceleration
errors at station $m$. The resulting residual round-trip phase is
\begin{equation}
\delta\phi_m(t)
=
\frac{4\pi}{\lambda}
\left(
\Delta v_mt+\frac{1}{2}\Delta a_mt^2
\right).
\label{eq:residual_phase}
\end{equation}
The fraction of coherent power retained over a CPI of duration $T_c$
is
\begin{equation}
\eta_m(T_c)
=
\left|
\frac{1}{T_c}
\int_0^{T_c}
e^{j\delta\phi_m(t)}\,dt
\right|^2.
\label{eq:coherent_loss}
\end{equation}
Accordingly, the actual coherently accumulated SNR is
\begin{equation}
\gamma_{c,m}(T_c)
=
\rho_mT_c\eta_m(T_c).
\label{eq:motion_limited_snr}
\end{equation}
For residual velocity alone,
\begin{equation}
\eta_m(T_c)
=
\sinc^2\!\left(
\frac{2\Delta v_mT_c}{\lambda}
\right).
\label{eq:velocity_coherent_loss}
\end{equation}
When residual acceleration is also present,
\eqref{eq:coherent_loss} is evaluated numerically.

For an allowed coherent-power retention $\eta_{\min}$, define the
residual-phase limit as the first duration satisfying
\begin{equation}
T_{\phi,m}
=
\inf
\left\{
T_c>0:
\eta_m(T_c)\leq\eta_{\min}
\right\}.
\label{eq:phase_coherence_limit}
\end{equation}
The numerical results use $\eta_{\min}=1/2$, corresponding to a
3~dB coherent-power loss.

Target translation can independently move the UAV echo across range
bins. Let $v_{r,m}$ and $a_{r,m}$ denote the uncompensated radial speed
and acceleration, respectively. Restricting the displacement to a
fraction $\xi_R$ of the range resolution
$\Delta R=c/(2B)$ gives
\begin{equation}
|v_{r,m}|T_c
+
\frac{1}{2}|a_{r,m}|T_c^2
\leq
\xi_R\frac{c}{2B}.
\label{eq:range_migration}
\end{equation}
For $|a_{r,m}|>0$, the corresponding range-migration limit is
\begin{equation}
T_{R,m}
=
\frac{
-|v_{r,m}|
+
\sqrt{
|v_{r,m}|^2
+
2|a_{r,m}|\xi_Rc/(2B)
}
}{
|a_{r,m}|
}.
\label{eq:range_migration_limit}
\end{equation}
When $a_{r,m}=0$, this reduces to
$T_{R,m}=\xi_Rc/(2B|v_{r,m}|)$.

The usable single-CPI duration is therefore bounded by
\begin{equation}
T_{{\rm coh},m}
=
\min\{T_{\phi,m},T_{R,m}\}.
\label{eq:coherence_time}
\end{equation}
An operating point is compatible with a single coherent CPI when the
required duration in \eqref{eq:total_observation_time} lies within this
limit and the loss in \eqref{eq:motion_limited_snr} is included in the
link margin. Otherwise, the body track must be used to update the
target delay and bulk Doppler over shorter CPIs. Combining such CPIs
over longer intervals is possible, but a detailed multi-CPI fusion
algorithm is beyond the scope of the present work.

Equations \eqref{eq:total_observation_time} and
\eqref{eq:coherence_time} jointly provide the design relationship
evaluated in Section~\ref{sec:results}: blade RCS, range, and transmit
power determine how much observation time is required, whereas
residual phase and range migration determine whether that duration can
be realized within one coherent processing interval.

\section{Numerical and Experimental Results}
\label{sec:results}

In this section, we validate the proposed OFDM-ISAC UAV estimation framework through extensive numerical simulations and hardware experiments. 

\subsection{Simulation Setup}

Unless otherwise stated, the
numerical evaluations use a carrier frequency of 28~GHz, a rotor radius
of 0.1~m, and a rotor rate of 60~revolutions/s. These parameters give a
blade-tip speed of 37.70~m/s and a Doppler scale of
$\kappa=7,042.0$~Hz. For the statistical multi-station evaluation, the three stations are
located at $[-45,-20,0]$, $[50,-15,5]$, and $[0,55,8]$~m, while the UAV
is located at $[8,12,65]$~m. A nominal ZYX attitude of
$(12^\circ,-18^\circ,35^\circ)$ is used to generate the true rotor-axis
direction. Independent bandwidth errors are generated according to
\eqref{eq:multiview_measurement}, and every Monte Carlo point is
averaged over 1,000 trials.  Different configurations
used by the signal-level simulation and hardware experiments are
specified in their corresponding subsections.

\begin{figure}[!t]
\centering
\subfloat[Required transmit power versus range for different blade-RCS values.]{
    \includegraphics[width=0.46\linewidth]
    {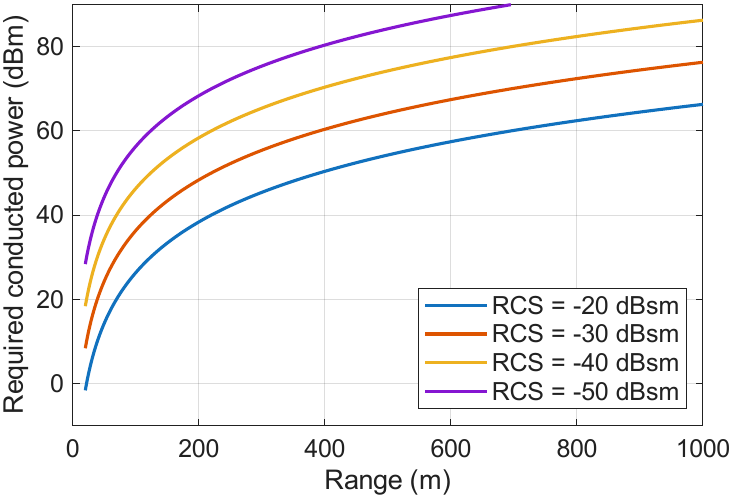}
}\hfill
\subfloat[Maximum sensing range versus transmit power for different sensing tasks.]{
    \includegraphics[width=0.46\linewidth]
    {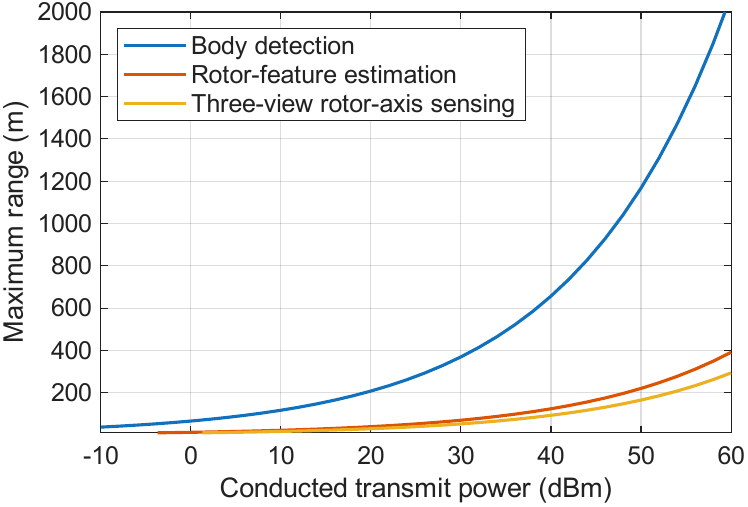}
}
\caption{Link-budget relationships among blade RCS, sensing range, and transmit power.}
\label{fig:link_budget}
\end{figure}

\subsection{Single-Station Rotor-Frequency Estimation}

We first examine the harmonic structure induced by rotor motion and
evaluate frequency-estimation performance under an ideal coherent
signal model. Fig.~\ref{fig:single_sim}(a) illustrates the spectral
harmonics generated by periodic rotation. For approximately identical
and uniformly spaced blades, blade symmetry favors harmonics at
integer multiples of the blade-pass frequency $B_qf_{r,q}$.

Fig.~\ref{fig:single_sim}(b) evaluates a separate weighted
harmonic-comb estimator applied directly to the complex
phase-modulated signal. The results are obtained with a slow-time sampling interval of $1$~ms, and $512$
samples, corresponding to an observation duration of $0.512$~s. The resulting rotor-frequency MSE is compared with the many-cycle
local CRB in \eqref{eq:frequency_crb}. At low SNR, incorrect frequency
selection produces large errors outside the regime described by the
local bound. As SNR increases, the MSE decreases toward the benchmark.
This comparison illustrates the performance of the weighted
harmonic-comb search.

\subsection{Multi-Station Rotor-Axis Estimation}

We next evaluate the performance of multi-station rotor-axis estimation. A
signal-level simulation first validates the complete processing pipeline
from blade echoes to bandwidth extraction and axis recovery. A separate
Monte Carlo simulation then evaluates the statistical accuracy of the
joint axis-and-scale estimator and compares it with the local CRB.

We first validate the complete processing pipeline
from blade echoes to bandwidth extraction and axis recovery. In this simulation, we consider three monostatic stations are located
at $[1,0,0]$, $[-1,0,0]$, and $[-1,-1,0]$~m, while the UAV center is
fixed at $[0,0,2]$~m as illustrated in Fig.~\ref{fig:multiview}(a). The carrier frequency is 5.8~GHz, and the
slow-time signal is sampled at 200~kHz over 3 seconds. The UAV contains four
two-blade rotors with a blade length of 7~cm and width of 2.5~cm. The
rotor rate is 100~revolutions/s, corresponding to a blade-tip speed of
43.98~m/s. Roll and yaw are fixed at zero, while the pitch varies as $\vartheta(t)
=
30^\circ\sin(2\pi\times1.6667t).$

\begin{figure}[!t]
\centering
\includegraphics[width=0.96\linewidth]
{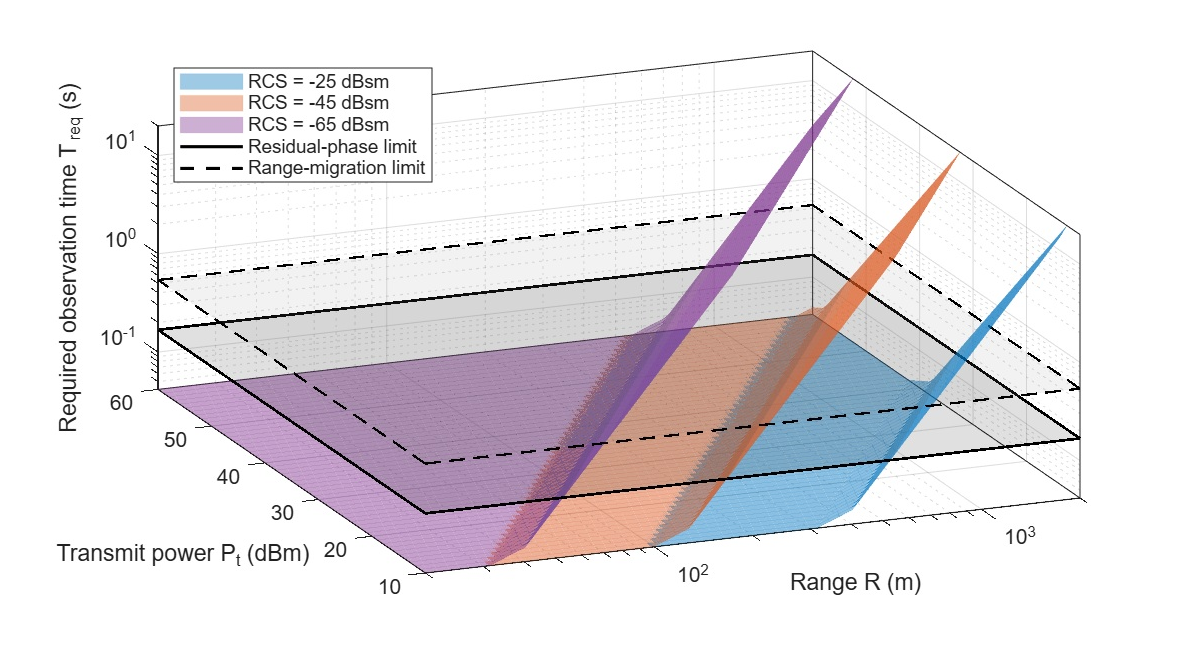}
\caption{Required observation time versus range and transmit power for
different blade-RCS values. The horizontal planes indicate the
residual-phase and range-migration limits.}
\label{fig:observation_time}
\end{figure}

Representative intermediate results are shown in
Fig.~\ref{fig:multiview_sim}. The time--velocity spectrum observed at
one station in Fig.~\ref{fig:multiview_sim}(a) exhibits the periodically
varying Doppler extent generated by the rotating blades. Because the
three stations observe different projections of the common rotor plane,
their extracted half-bandwidths follow distinct trajectories, as shown
in Fig.~\ref{fig:multiview_sim}(b). These complementary measurements
are then fused to recover the time-varying rotor axis. The estimated pitch in Fig.~\ref{fig:multiview_sim}(c) follows the
amplitude and periodicity of the imposed motion, with an RMSE of
$8.68^\circ$. The independently estimated rotor rate is
99.56~revolutions/s, close to the prescribed value of
100~revolutions/s. The remaining pitch error is mainly caused by finite
Doppler resolution, aspect-dependent blade returns, bandwidth-extraction
errors, and locally unfavorable viewing geometry. These results
validate the complete signal-level multi-view processing chain.

We further evaluate the statistical accuracy of the rotor-axis
estimator and compare it with the theoretical CRB. Fig.~\ref{fig:multiview}(b) presents the
acute rotor-axis angular RMSE as the standard deviation of the
per-station bandwidth error increases. The proposed joint estimator
treats both $\vect a$ and $\kappa$ as unknown, whereas the benchmark
fixes $\kappa$ to its true value. Each estimator is compared with its
corresponding local CRB. We can observe that at low bandwidth uncertainty, the simulated RMSE closely follows the
CRB, validating the Jacobian and Fisher-information analysis in
Section~\ref{sec:multi}. As the uncertainty increases, nonlinear
curvature, the unit-sphere constraint, and estimator bias cause the
RMSE to depart from the local bound. Fixing $\kappa$ improves the
accuracy because it removes the tip-speed scale as a nuisance parameter.
Nevertheless, joint estimation remains necessary when the effective
blade radius or tip speed is not known in advance.

\subsection{Sensing Range and Power Requirements}
In this subsection, we evaluate how blade RCS, sensing range, and transmit power jointly determine whether the weak micro-Doppler return can achieve the required post-processing SNR. The link-feasibility evaluation uses transmit and receive antenna gains
of 25~dBi, a receiver noise figure of 7~dB, a system loss of 3~dB, and
an equivalent noise bandwidth of 25~MHz. A 128-sample coherent
processing gain, corresponding to 21.07~dB, is assumed. The illustrative
task parameters are a body RCS of $-10$~dBsm with a 6~dB detection
threshold, and a blade RCS of $-35$~dBsm with thresholds of 10~dB for
rotor-feature extraction and 15~dB for a usable rotor-axis link. These
values are intended to expose the relative link requirements rather
than establish universal UAV sensing limits.

Fig.~\ref{fig:rcs_aspect} evaluates the rectangular-blade model in
\eqref{eq:blade_rcs} for a 7~cm by 2.5~cm blade. The ideal
perfect-conductor result changes sharply with incidence angle, showing
that a single aspect-independent blade RCS can be misleading. The
second curve introduces a representative $-20$~dB material-efficiency
factor. Its purpose is to demonstrate the sensitivity of the predicted
link budget to blade material and modeling assumptions; quantitative
range claims therefore require experimentally calibrated blade RCS.

\begin{figure}[!t]
\centering
\subfloat[Single-station ISAC hardware.]{
    \includegraphics[width=0.45\linewidth]{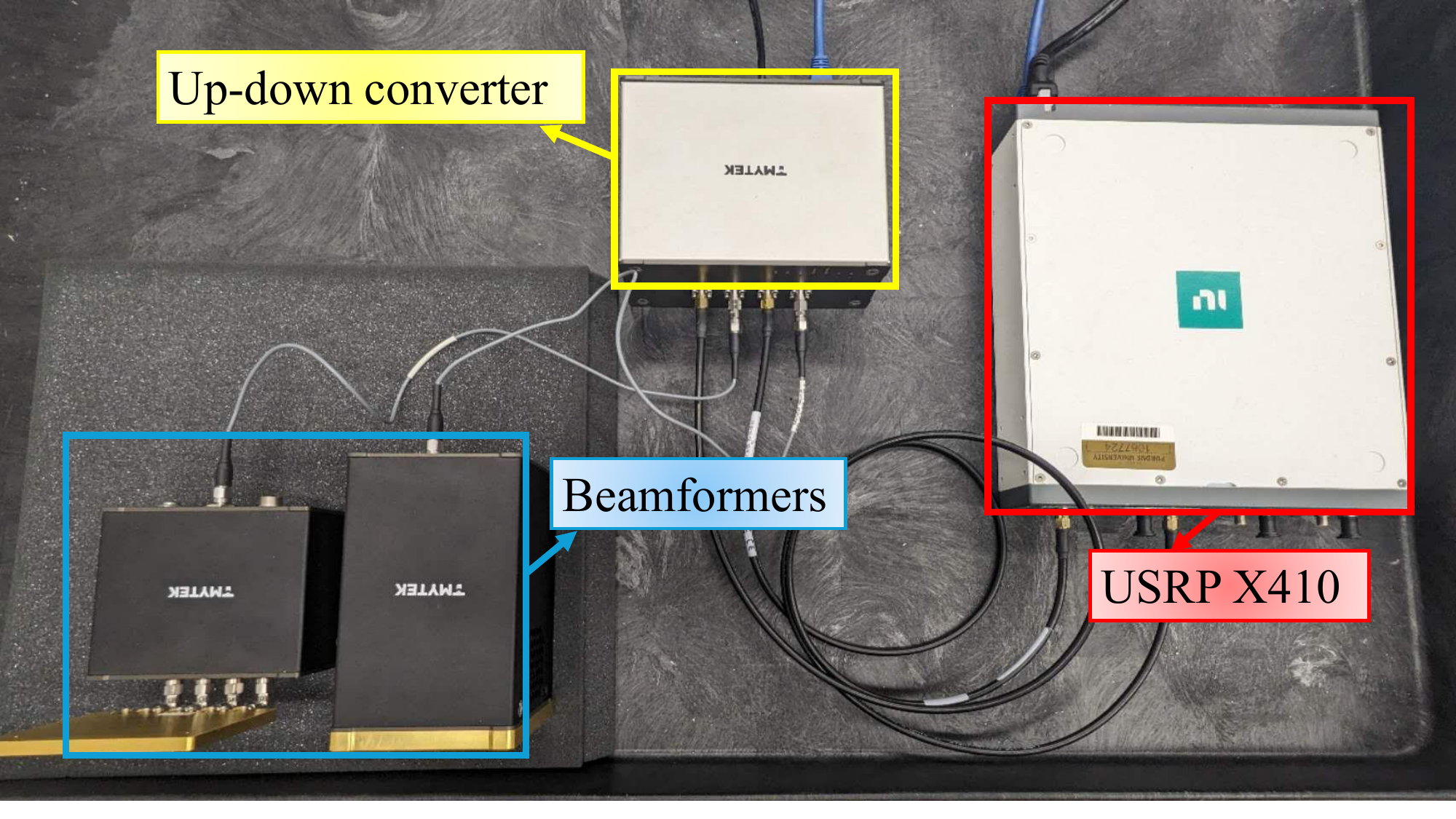}
}\hfill
\subfloat[Single-station measurement scene.]{
    \includegraphics[width=0.45\linewidth]{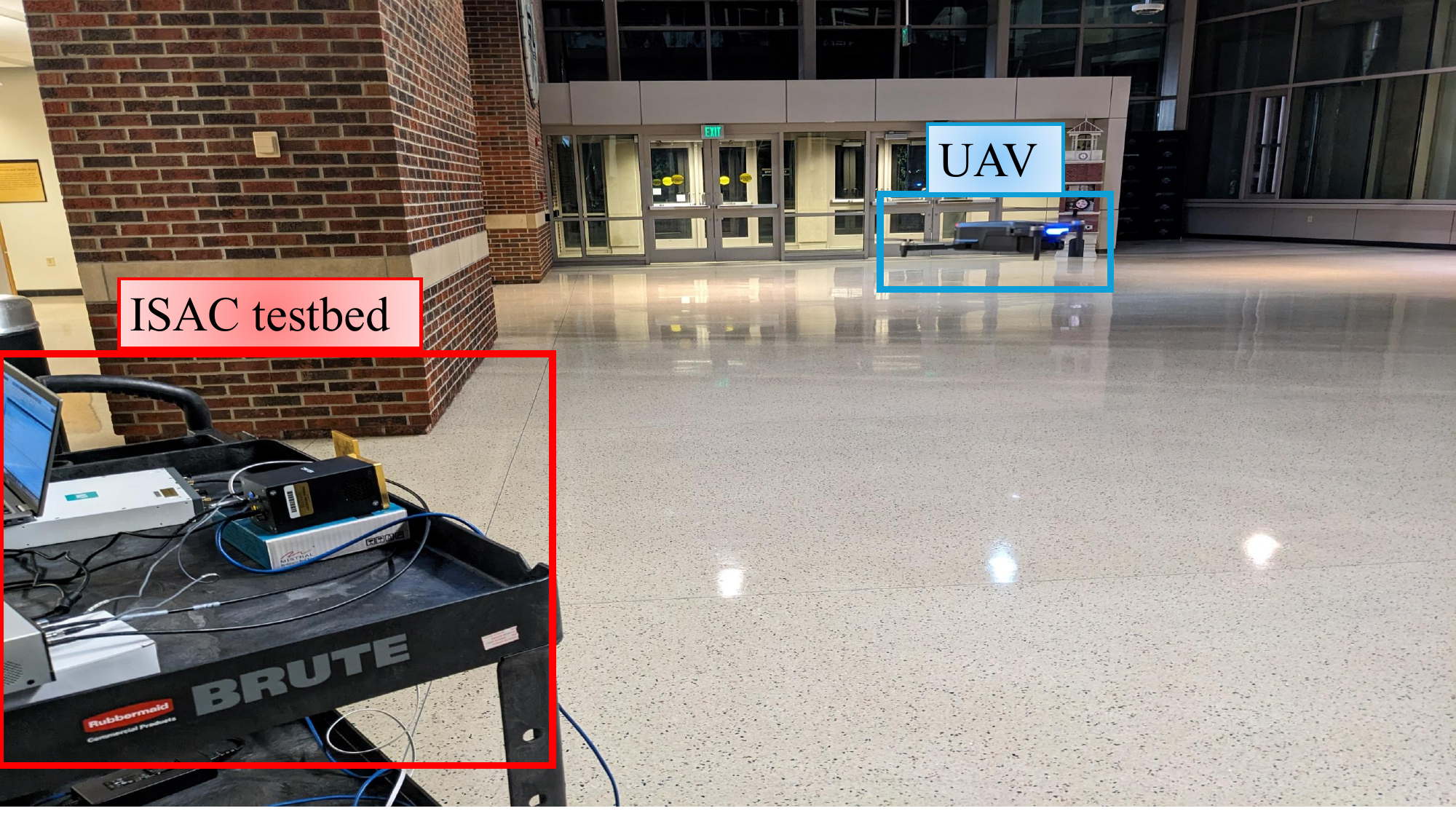}
}\\
\subfloat[Three-station ISAC hardware.]{
    \includegraphics[width=0.45\linewidth]{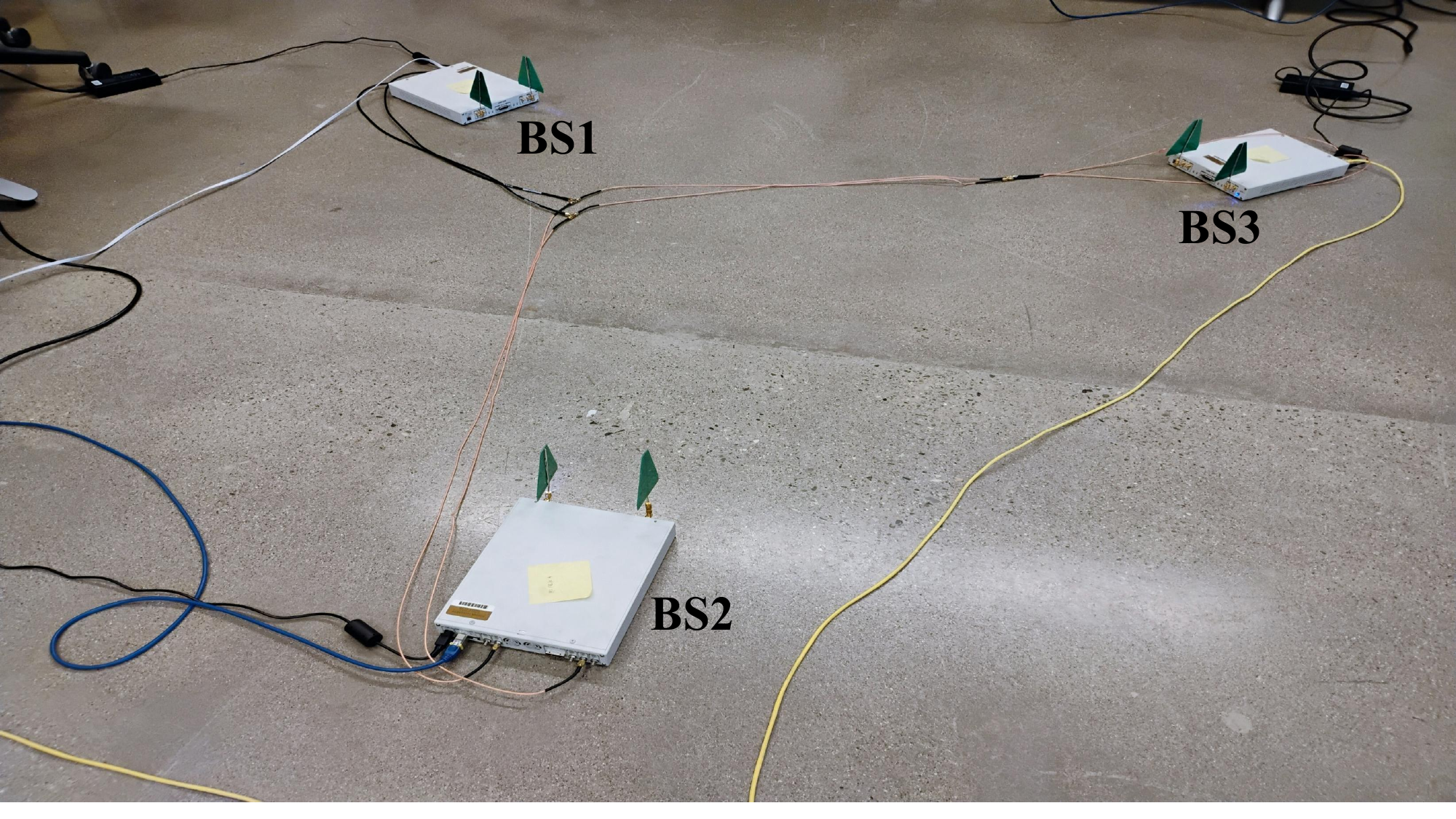}
}\hfill
\subfloat[Three-station measurement scene.]{
    \includegraphics[width=0.45\linewidth]{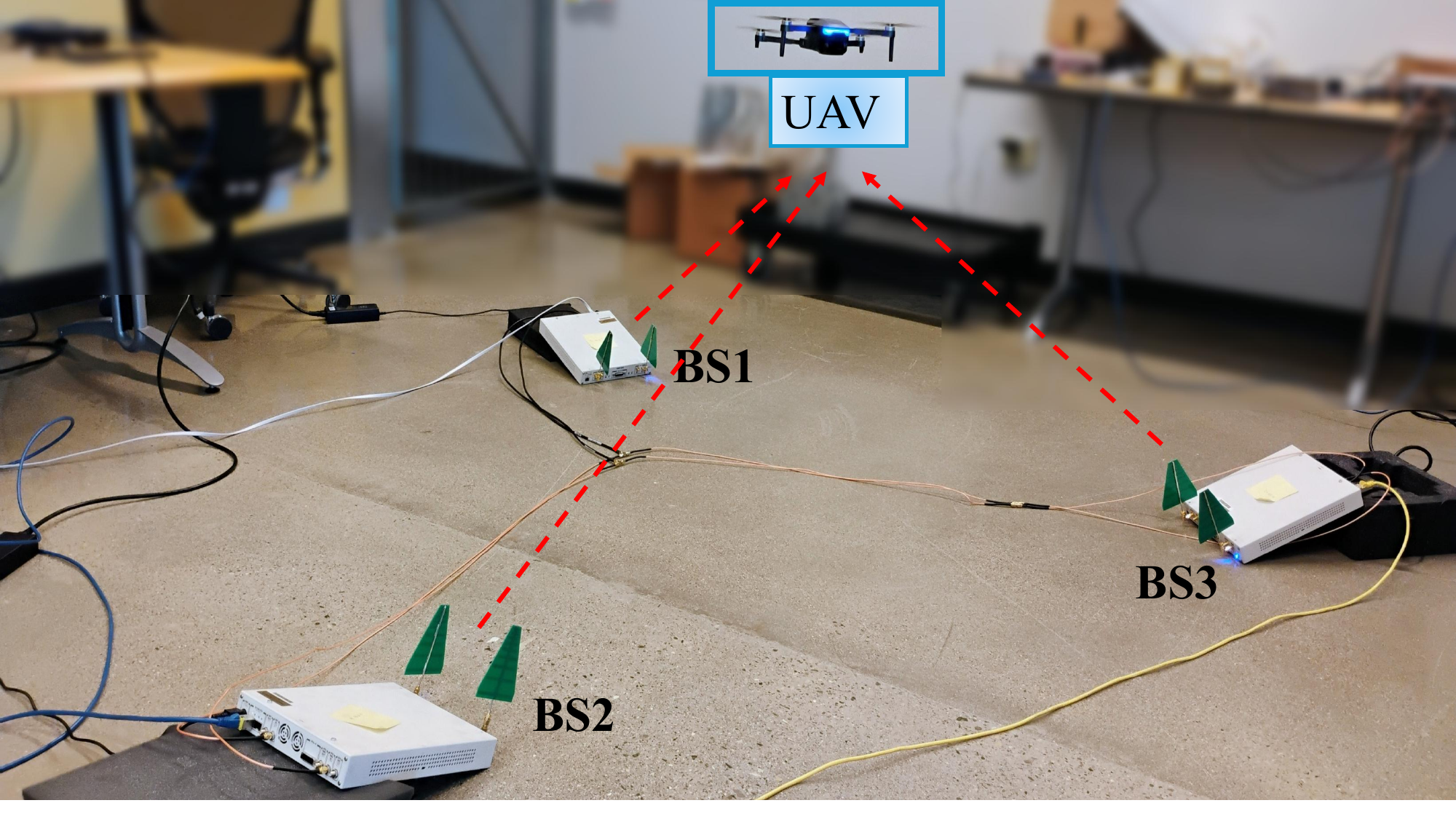}
}
\caption{Hardware platforms and measurement environments used for the
single-station and multi-station experiments.}
\label{fig:experiment_setup}
\end{figure}

Fig.~\ref{fig:link_budget}(a) shows the required transmit power as a
function of sensing range for different blade-RCS values, whereas
Fig.~\ref{fig:link_budget}(b) presents the corresponding maximum sensing
range versus transmit power for body detection, rotor-feature detection,
and rotor-axis estimation. As expected from the monostatic $R^4$
propagation law, the required transmit power increases rapidly with
range, while a lower blade RCS produces a comparable increase in the
required link budget. The feature-dependent curves further show that
reliable UAV-body detection does not necessarily guarantee successful
rotor sensing. Rotor-feature and rotor-axis estimation impose more
stringent link requirements because the blade return is weaker and a
higher post-processing SNR is needed for reliable micro-Doppler
bandwidth extraction. Consequently, their achievable sensing ranges are
substantially shorter than that of body detection at the same transmit
power. Moreover, because every station participating in the
multi-station estimator must provide a usable bandwidth measurement,
the weakest participating link can limit the common sensing range.

\subsection{Observation Time and Motion Limitations}

In this subsection, we further evaluate the observation time required for reliable
micro-Doppler feature extraction.

Fig.~\ref{fig:observation_time} presents a three-dimensional design
chart in which the required observation time is plotted against sensing
range and conducted transmit power. The three surfaces correspond to
blade RCS values of $-25$, $-45$, and $-65$~dBsm, while the horizontal
planes indicate the residual-phase and range-migration limits. The
observation time at each operating point is calculated from
\eqref{eq:total_observation_time} as the maximum of the durations
required to satisfy a 10~dB feature-detection threshold, a 1~Hz
rotor-rate RMSE with an effective phase-modulation index $\beta=3$, and
five blade-pass cycles for a two-blade rotor rotating at
60~revolutions/s.

\begin{figure}[!t]
\centering
\subfloat[Heatmap before MTI.]{
    \includegraphics[width=0.44\linewidth]
    {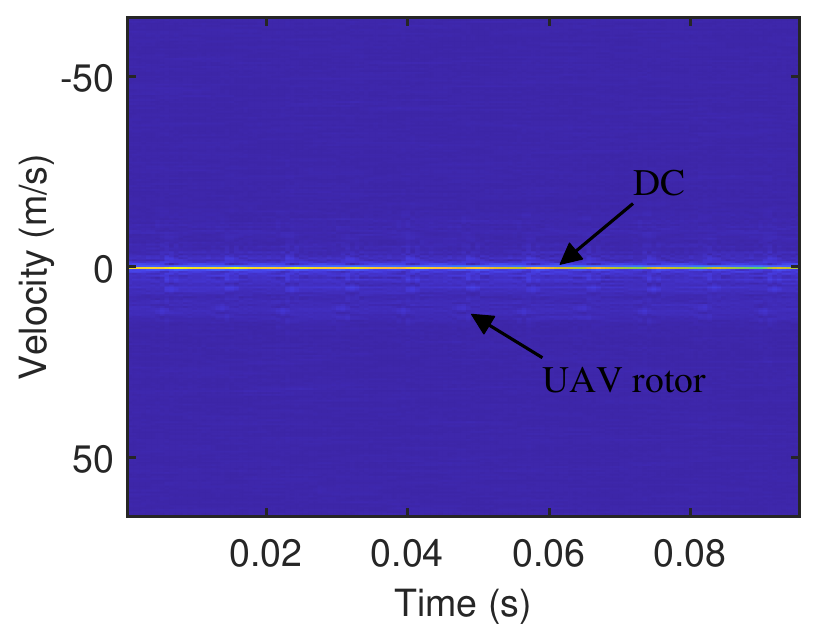}
}\hfill
\subfloat[Heatmap after MTI.]{
    \includegraphics[width=0.44\linewidth]
    {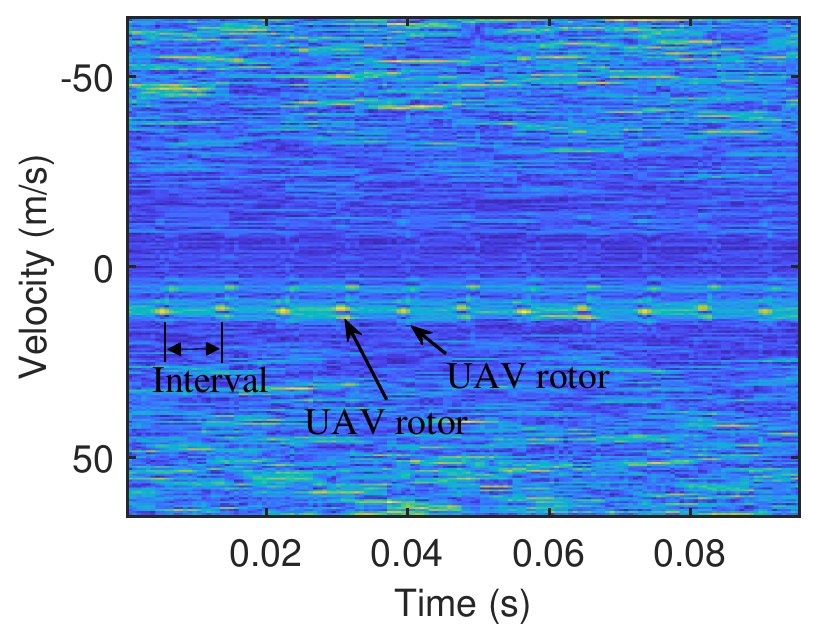}
}\\
\subfloat[Measured Doppler spectrum.]{
    \includegraphics[width=0.68\linewidth]
    {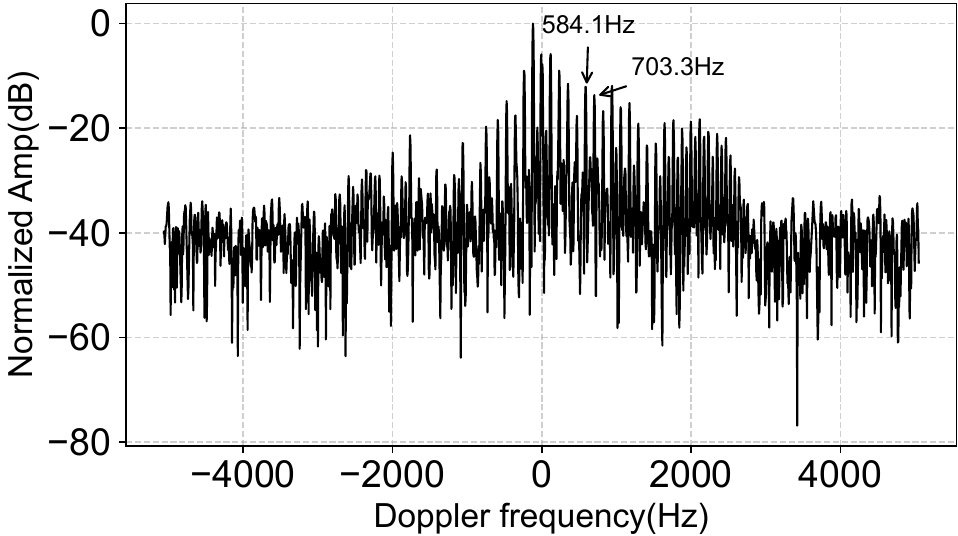}
}
\caption{Measured single-station micro-Doppler signatures before and
after MTI and the resulting Doppler spectrum.}
\label{fig:experiment_signal}
\end{figure}

At short range or high transmit power, the required observation time
approaches the 41.7~ms cycle-limited floor because a minimum number of
blade-pass cycles must be collected. In this regime, increasing transmit
power provides little further reduction in observation time. As the
range increases or the blade RCS decreases, the blade-return energy
becomes the dominant limitation, and the required duration increases
rapidly according to the $R^4/(P_t\sigma_b)$ scaling. Consequently,
increasing transmit power or blade RCS shifts the required-time surface
downward, whereas increasing range shifts it upward. These trends also
show that transmit power and blade RCS play equivalent roles in the
energy-limited regime through their product $P_t\sigma_b$.

\begin{figure*}[t]
\centering
\subfloat[Heatmap before MTI.]{
    \includegraphics[width=0.22\textwidth]{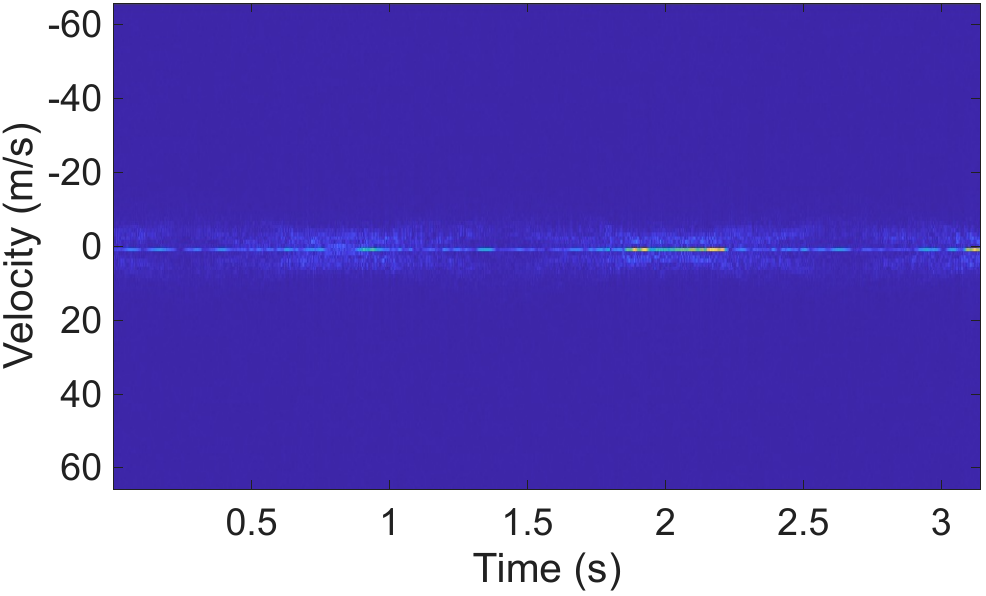}
}\hfill
\subfloat[Heatmap after MTI.]{
    \includegraphics[width=0.22\textwidth]{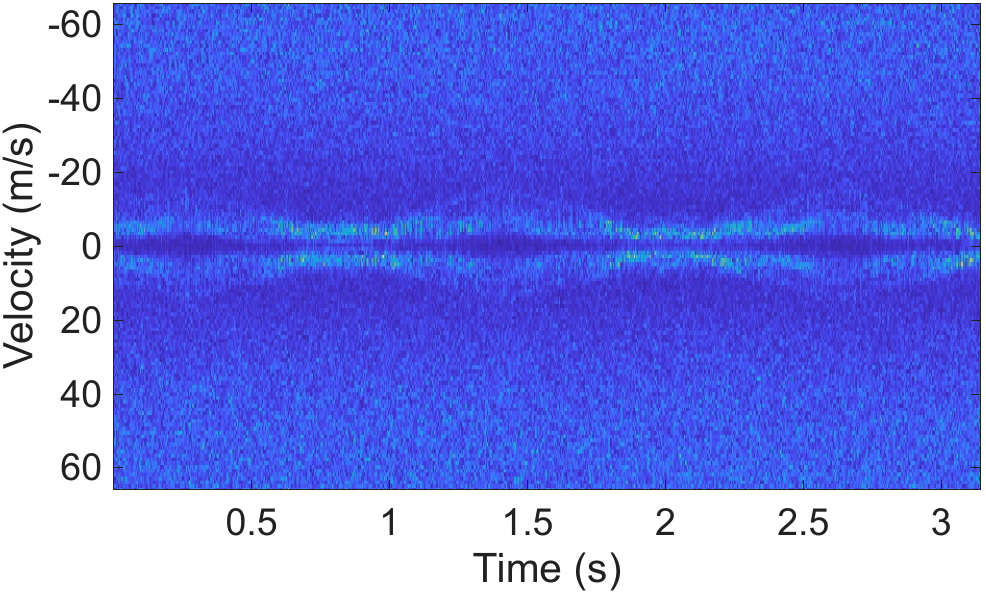}
}\hfill
\subfloat[Micro-Doppler bandwidth envelope.]{
    \includegraphics[width=0.22\textwidth]{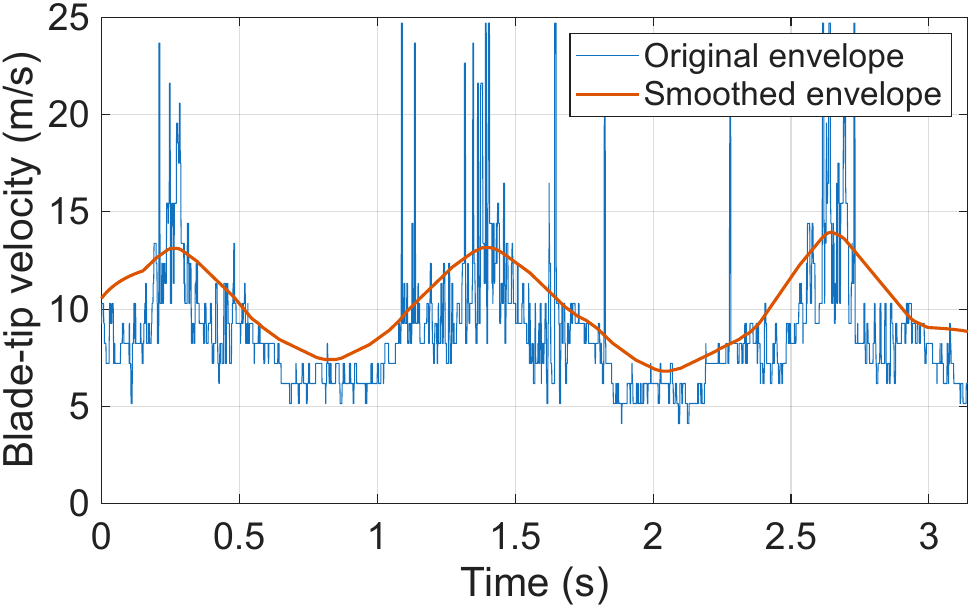}
}\hfill
\subfloat[Estimated roll-related axis angle.]{
    \includegraphics[width=0.22\textwidth]{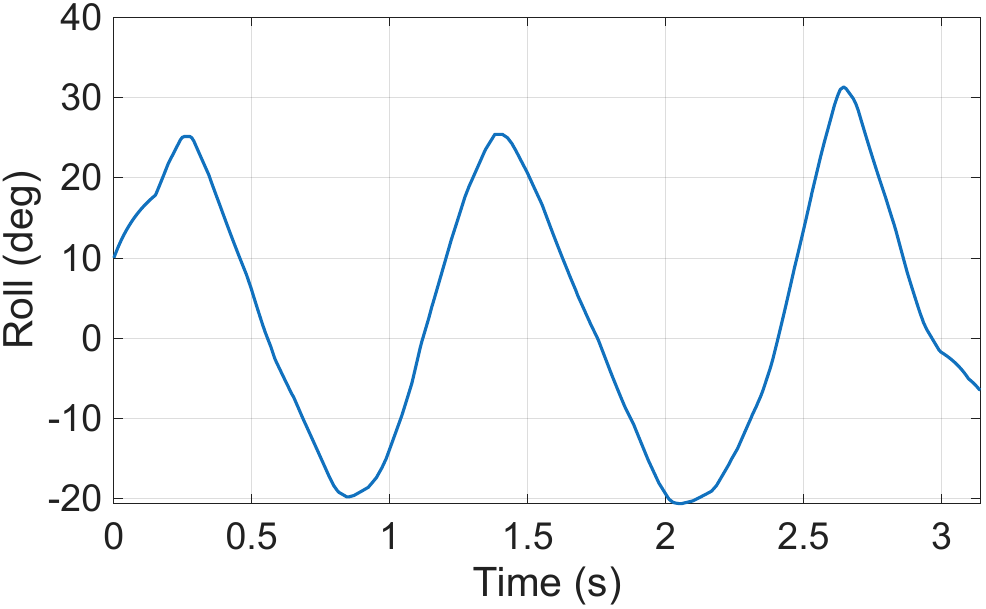}
}
\caption{Experimental multi-station rotor-axis estimation and
representative intermediate processing results from one station.}
\label{fig:multiview_exp}
\end{figure*}

\begin{figure*}[!t]
\centering
\subfloat[Measured configuration 1.]{
    \includegraphics[width=0.185\linewidth]{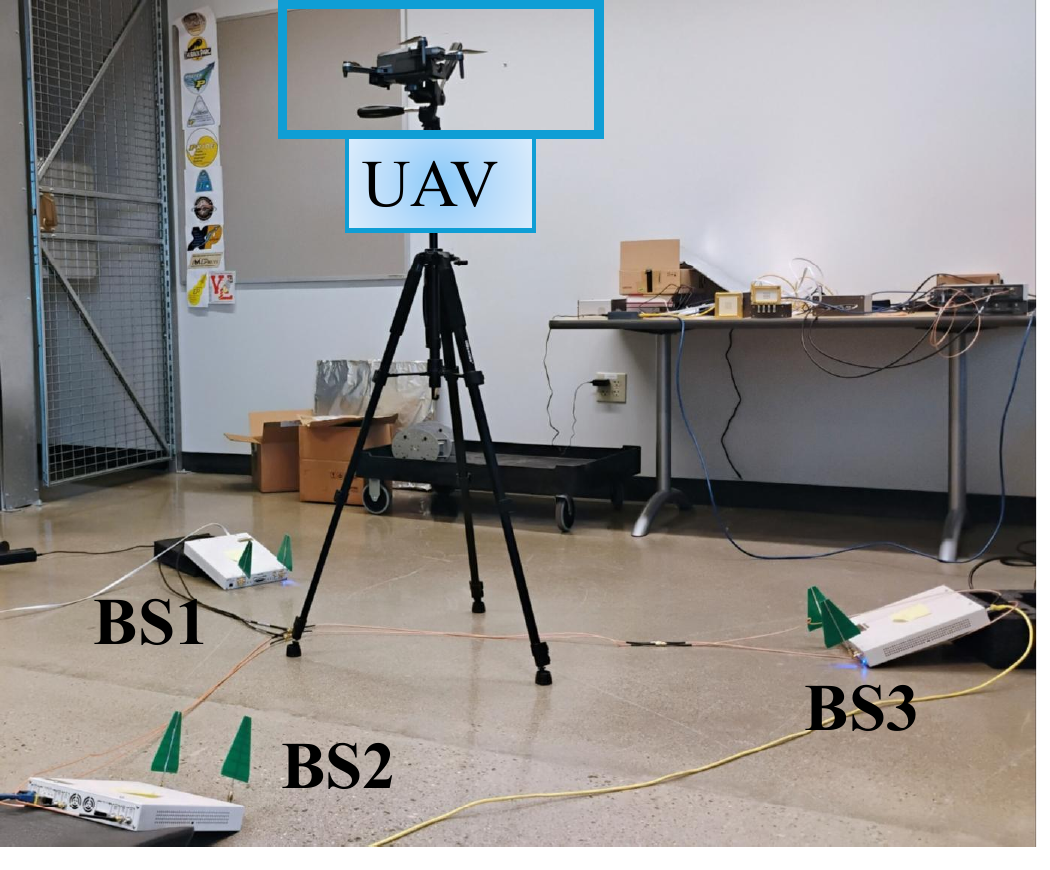}
}\hfill
\subfloat[Measured configuration 2.]{
    \includegraphics[width=0.185\linewidth]{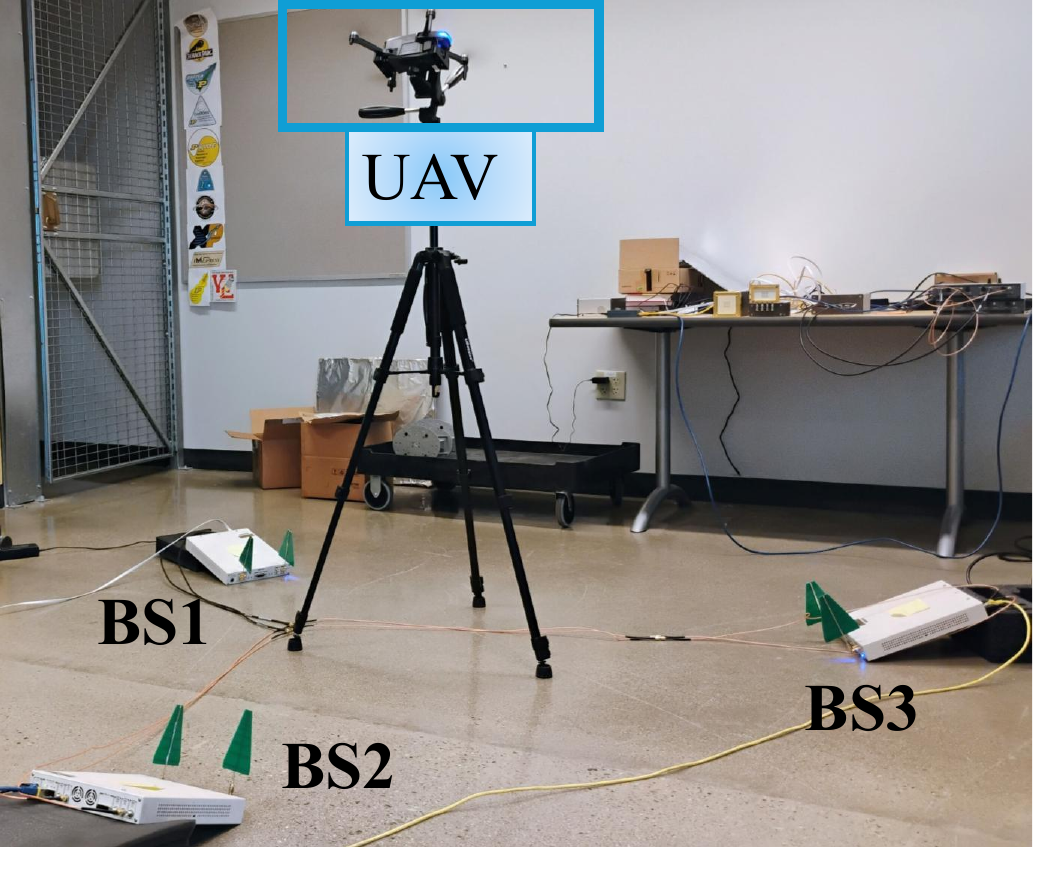}
}\hfill
\subfloat[Measured configuration 3.]{
    \includegraphics[width=0.185\linewidth]{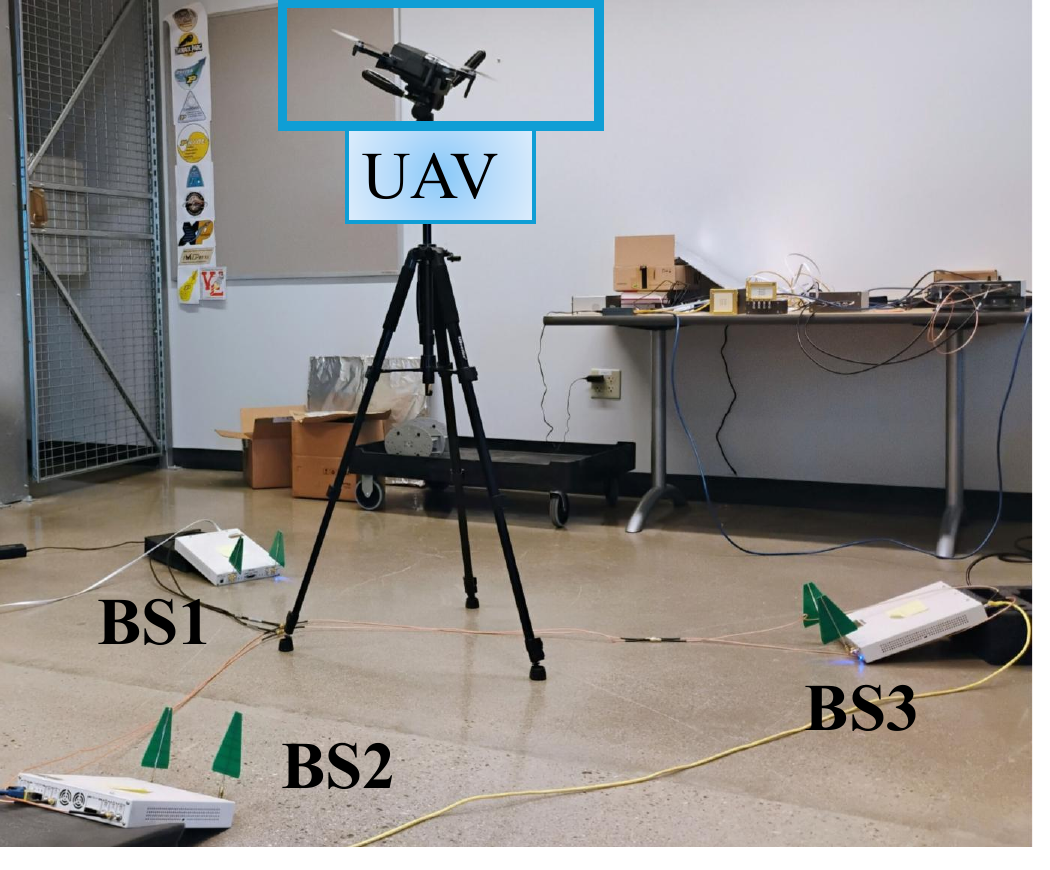}
}\hfill
\subfloat[Measured configuration 4.]{
    \includegraphics[width=0.185\linewidth]{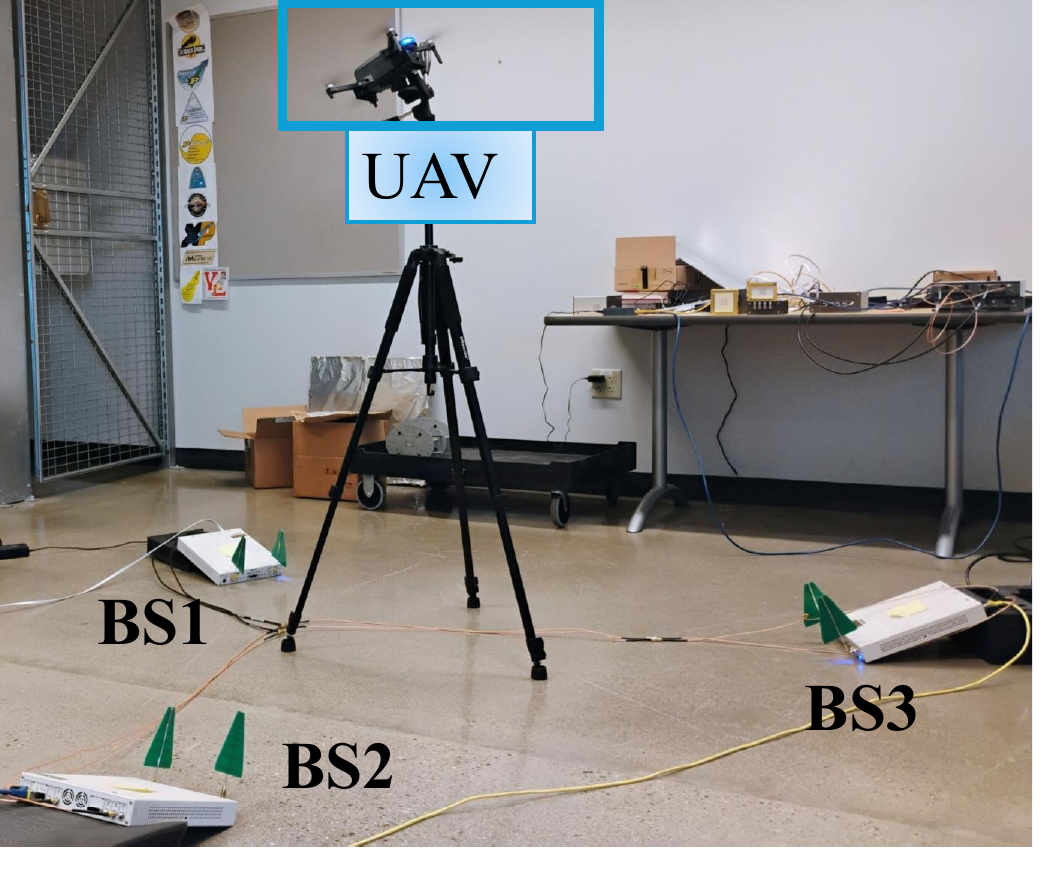}
}\hfill
\subfloat[Measured configuration 5.]{
    \includegraphics[width=0.185\linewidth]{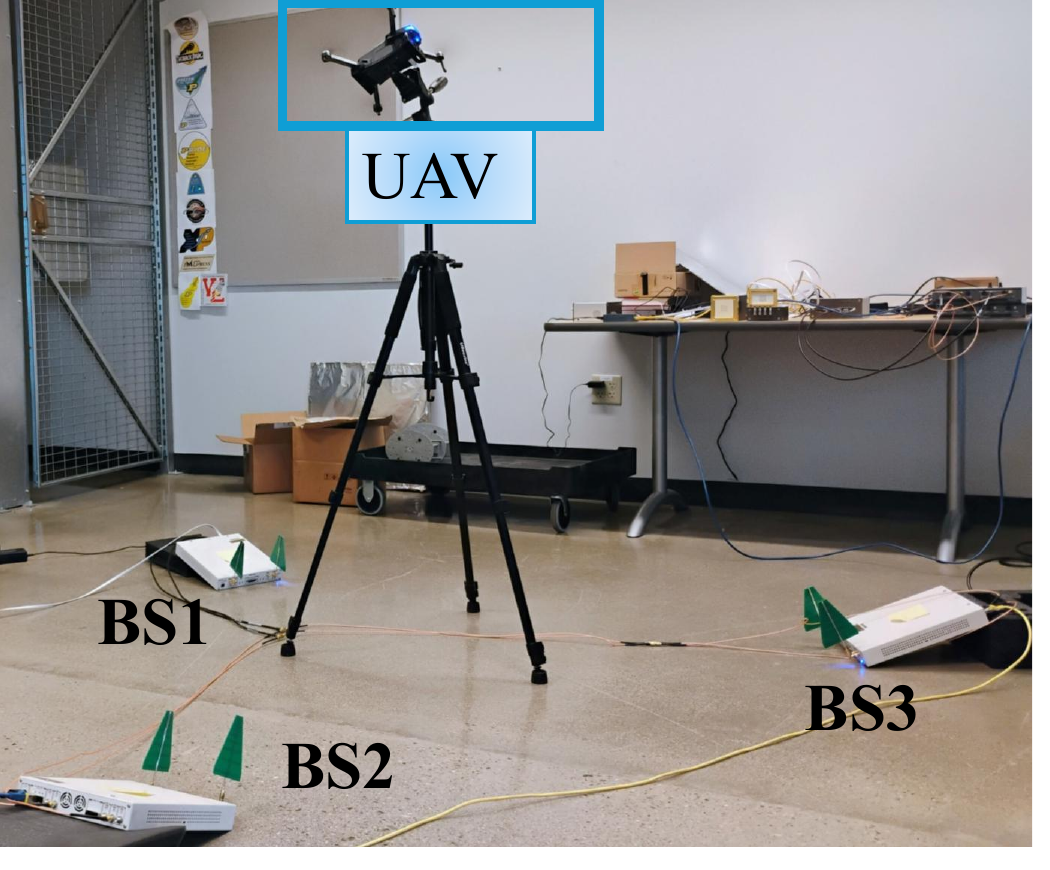}
}\\
\subfloat[Estimated axis 1.]{
    \includegraphics[width=0.185\linewidth]{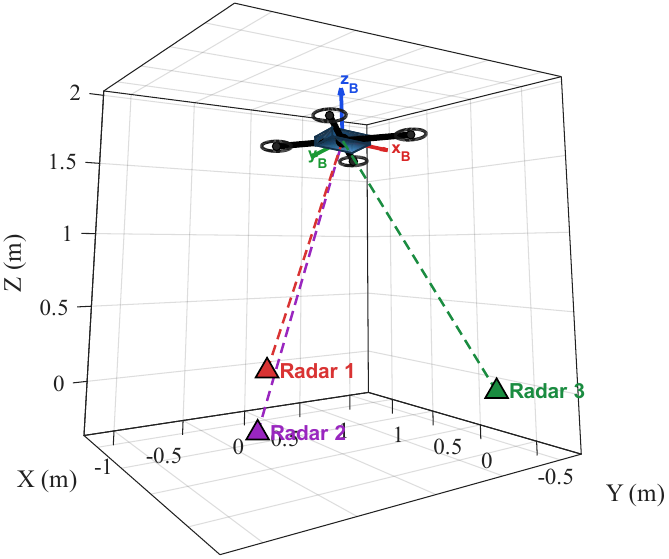}
}\hfill
\subfloat[Estimated axis 2.]{
    \includegraphics[width=0.185\linewidth]{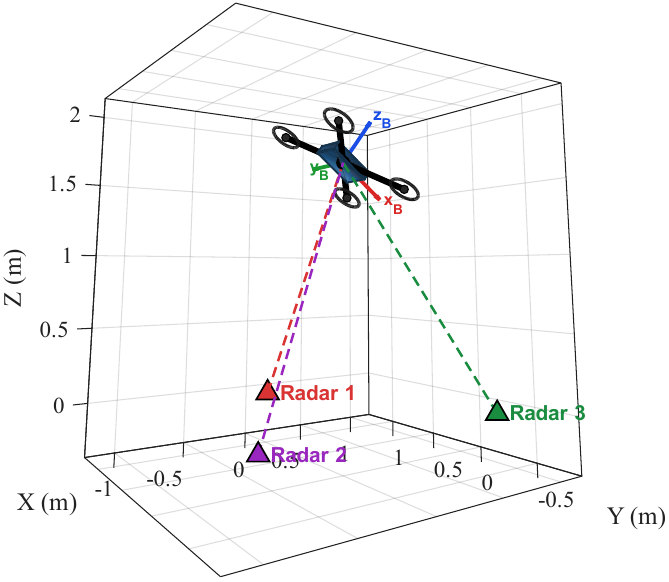}
}\hfill
\subfloat[Estimated axis 3.]{
    \includegraphics[width=0.185\linewidth]{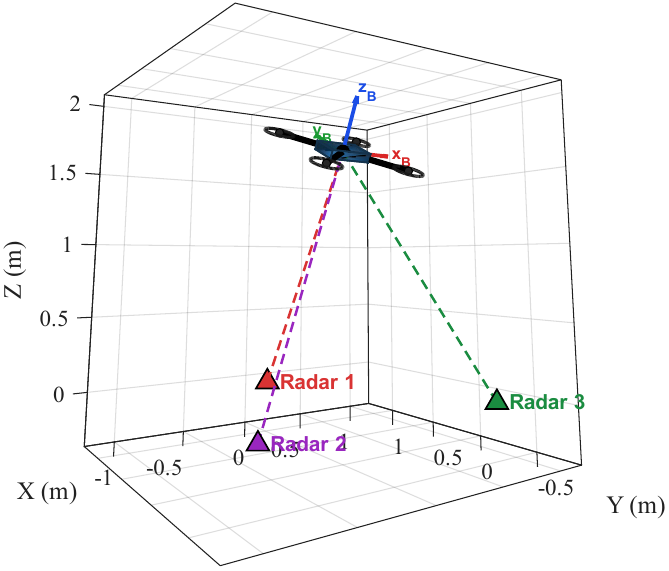}
}\hfill
\subfloat[Estimated axis 4.]{
    \includegraphics[width=0.185\linewidth]{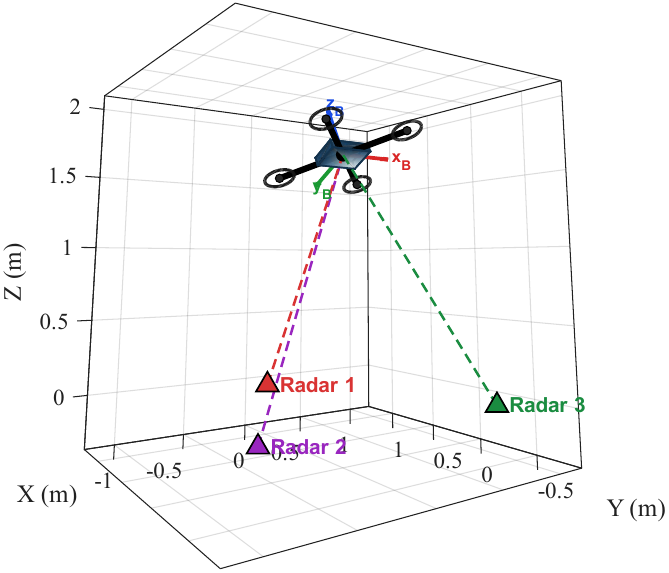}
}\hfill
\subfloat[Estimated axis 5.]{
    \includegraphics[width=0.185\linewidth]{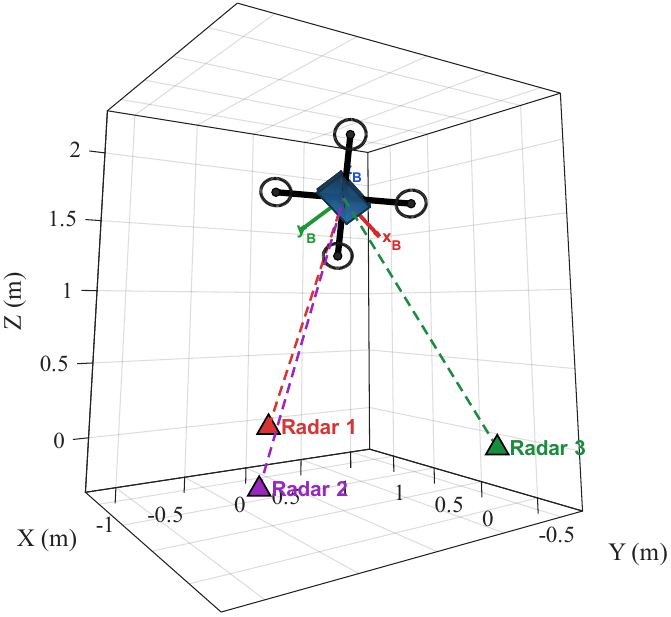}
}
\caption{Multi-station experimental results for five manually
configured UAV orientations. The top row shows the measured
configurations, and the bottom row shows the corresponding estimated
rotor-axis geometries.}
\label{fig:experiment_results}
\end{figure*}

The horizontal planes indicate whether the required observation can be
supported within a single coherent processing interval. One plane
represents the residual-phase coherence limit, while the other represents
the range-migration limit under the assumed motion conditions. Operating
points well below both limits can be processed within one CPI, whereas
points approaching or exceeding either boundary require improved motion
compensation or integration over multiple shorter CPIs. 

\subsection{Experimental Validation}

We finally use our ISAC prototype to validate the feasibility of the proposed system. The experimental platforms and measurement scenes are shown in
Fig.~\ref{fig:experiment_setup}. The single-station experiment uses an
NI USRP X410, external up/down converters, and analog transmit and
receive beamformers operating at 28~GHz. The multi-station experiment
uses three USRP X310 devices operating at different central frequencies separately to avoid inter-interference.

\subsubsection{Single-Station Rotor-Rate Estimation}

The single-station OFDM waveform contains 1024 subcarriers over a
25~MHz bandwidth. The slow-time processing uses 128-symbol STFT windows
with a 16-symbol hop. The UAV hovers in the atrium shown in
Fig.~\ref{fig:experiment_setup}(b), allowing the range-focused body and
blade returns to be observed within the same range cell.

Before MTI, the heatmap in Fig.~\ref{fig:experiment_signal}(a) is dominated by the
strong static and UAV-body component near zero Doppler. After MTI,
Fig.~\ref{fig:experiment_signal}(b) reveals the weaker periodic rotor
signature. Approximately 12 blade passages occur within 0.1~s. For the
two-blade rotor, this corresponds to a rotation rate close to 60~Hz.
The measured spectrum in Fig.~\ref{fig:experiment_signal}(c) contains
the corresponding harmonic structure, and the rotor-rate estimator
returns approximately 59.6~Hz. The agreement among the time--frequency
periodicity, spectral spacing, and estimated rate demonstrates
single-station rotor sensing.

\subsubsection{Multi-Station Rotor-Axis Estimation}

The multi-station experiment uses three distributed sensing stations,
each comprising one USRP X310 with separate log-periodic directional
antennas for transmission and reception. A common 10~MHz reference and
PPS signal are distributed by cable to synchronize the stations in
frequency and time. The devices are started using PPS-referenced timed
commands rather than a common sample-level trigger, so small
frame-level timing offsets can remain. Such offsets are acceptable
because the proposed method fuses time-aligned bandwidth features and
does not require carrier-phase coherence among the stations.

Because the experiment does not employ power amplifiers or low-noise
amplifiers, it is conducted over a short range in a controlled
laboratory environment. The UAV attitude is manually varied to produce
repeatable multi-view measurements. The purpose of this experiment is
to demonstrate feature extraction and rotor-axis fusion and long-range
hardware validation is left for future work.

Figs.~\ref{fig:multiview_exp}(a) and
\ref{fig:multiview_exp}(b) show the time--frequency spectrum at one
representative station before and after MTI. The strong component near
zero Doppler is substantially suppressed, exposing the rotor-induced
spectral extent. The extracted half-bandwidth in
Fig.~\ref{fig:multiview_exp}(c) supplies the local measurement used by
the multi-station estimator. After fusing the bandwidths from all three
stations, the recovered roll-related axis angle in
Fig.~\ref{fig:multiview_exp}(d) follows the manually imposed variation.
This result demonstrates that the view-dependent micro-Doppler
bandwidths can be combined to recover changes in the observable rotor
axis.

We further visualize the rotor-axis estimates for five different UAV
orientations. The UAV is mounted on a tripod so that its tilt can be
manually adjusted while its position remains fixed and is treated as
known. In a practical deployment, the UAV position can instead be
estimated by multi-station triangulation or by angle-of-arrival
estimation when antenna arrays are available. The manually configured
orientations are shown in Figs.~\ref{fig:experiment_results}(a)--(e),
and the corresponding estimated geometries are presented in
Figs.~\ref{fig:experiment_results}(f)--(j). The top row shows the
measurement configurations, while the bottom row visualizes the
rotor-axis directions recovered by fusing the three bandwidth
measurements. The estimated axes consistently reproduce the principal
changes in the imposed UAV tilt across the five configurations,
demonstrating the feasibility of the proposed multi-station
micro-Doppler approach. 

\section{Conclusion}
\label{sec:conclusion}

This paper developed a multi-view OFDM-ISAC framework for UAV rotor-rate
and rotor-axis sensing using micro-Doppler signatures. A single sensing
station estimates the rotor rate and view-dependent micro-Doppler
bandwidth, while multiple geometrically diverse stations fuse these
measurements to recover the common rotor-axis direction and Doppler
scale. A feature-aware feasibility analysis further connected blade RCS,
sensing range, transmit power, and observation duration while accounting
for motion-induced coherence and range-migration constraints. Numerical and experimental results validated the proposed single- and multi-station processing pipelines and demonstrated that, beyond conventional ISAC sensing of UAV position and bulk velocity, the proposed framework can extract additional rotor-level dynamic and orientation information for richer characterization of UAV operational states.

\bibliographystyle{IEEEtran}
\bibliography{references}

@article{liu2022isac,
  author={Fan Liu and Yuanhao Cui and Christos Masouros and Jie Xu and Tony Xiao Han and Yonina C. Eldar and Stefano Buzzi},
  title={Integrated Sensing and Communications: Toward Dual-Functional Wireless Networks for {6G} and Beyond},
  journal={IEEE Journal on Selected Areas in Communications},
  volume={40},
  number={6},
  pages={1728--1767},
  year={2022},
  month={Mar.},
  doi={10.1109/JSAC.2022.3156632}
}

@article{meng2024uav,
  author={Kezhi Meng and Qingqing Wu and Jie Xu and Wen Chen and Zhiyong Feng and Robert Schober and A. Lee Swindlehurst},
  title={{UAV}-Enabled Integrated Sensing and Communication: Opportunities and Challenges},
  journal={IEEE Wireless Communications},
  volume={31},
  number={2},
  pages={97--104},
  year={2024},
  month={Apr.},
}

@article{mu2023uav,
  author={Jianfeng Mu and Rui Zhang and Yuanhao Cui and Na Gao and Xiaojun Jing},
  title={{UAV} Meets Integrated Sensing and Communication: Challenges and Future Directions},
  journal={IEEE Communications Magazine},
  volume={61},
  number={5},
  pages={62--67},
  year={2023},
  month={Jan.},
}

@article{wei2025rotor,
  author={Jiachen Wei and Dingyou Ma and Feiyang He and Qixun Zhang and Zhiyong Feng and Zhengfeng Liu and Taohong Liang},
  title={{UAV}'s Rotor Micro-{D}oppler Feature Extraction Using Integrated Sensing and Communication Signal: Algorithm Design and Testbed Evaluation},
  journal={IEEE Transactions on Wireless Communications},
  volume={24},
  number={12},
  pages={10166--10182},
  year={2025},
  month={Jul.},
  doi={10.1109/TWC.2025.3578033}
}

@inproceedings{costa2024distributed,
  author={Heraldo Cesar Alves Costa and Saw James Myint and Carsten Andrich and Sebastian W. Giehl and Christian Schneider and Reiner S. Thom\"a},
  title={Modelling Micro-{D}oppler Signature of Drone Propellers in Distributed {ISAC}},
  booktitle={IEEE Radar Conference (RadarConf)},
  pages={472--477},
  year={2024},
  month={May},
  address={Denvor, CO, USA},
  doi={10.1109/RadarConf2458775.2024.10548468}
}

@INPROCEEDINGS{lu2024multibs,
  author={Lu, Xi and Wei, Zhiqing and Xu, Ruizhong and Wang, Lin and Lu, Bohao and Piao, Jinghui},
  booktitle={IEEE International Conference on Communications Workshops (ICC Workshops)}, 
  title={Integrated Sensing and Communication Enabled Multiple Base Stations Cooperative {UAV} Detection}, 
  year={2024},
  month={Jun.},
  address={Denver, CO, USA},
  volume={},
  number={},}

@article{zhang2019resolution,
  author={Hongbo Sun and Beom-Seok Oh and Xin Guo and Zhiping Lin},
  title={Improving the {D}oppler Resolution of Ground-Based Surveillance Radar for Drone Detection},
  journal={IEEE Transactions on Aerospace and Electronic Systems},
  volume={55},
  number={6},
  pages={3667--3673},
  year={2019},
  month = {Jan.},
  doi={10.1109/TAES.2019.2895585}
}

@book{chen2019microdoppler,
  author={Victor C. Chen},
  title={The Micro-Doppler Effect in Radar},
  edition={2},
  publisher={Artech House},
  year={2019}
}

@INPROCEEDINGS{pu2026conference,
  author={Pu, Henglin and Su, Lu and Li, Husheng},
  booktitle={IEEE International Conference on Communications (ICC)}, 
  title={{ISAC} Micro-{D}oppler Sensing of {UAV}s: Cramér-Rao Bound Analysis and Experiment Demonstration}, 
  year={2026},
  month = {May},
  address={Glasgow, UK},
  volume={},
  number={},
  doi={10.1109/ICC59461.2026.11587456}}

@article{park2021antidrone,
  author={Seongjoon Park and Hyeongtaek T. Kim and Sangmin Lee and Hyeontae Joo and Hwangnam Kim},
  title={Survey on Anti-Drone Systems: Components, Designs, and Challenges},
  journal={IEEE Access},
  volume={9},
  pages={42635--42659},
  year={2021},
  month={Mar.},
  doi={10.1109/ACCESS.2021.3065926}
}

@article{patel2018review,
  author={Jash J. Patel and Francesco Fioranelli and David Anderson},
  title={Review of Radar Classification and {RCS} Characterisation Techniques for Small {UAV}s or Drones},
  journal={IET Radar, Sonar \& Navigation},
  volume={12},
  number={9},
  pages={911--919},
  year={2018},
  month={Jul.},
  doi={10.1049/iet-rsn.2018.0020}
}

@article{chen2006microdoppler,
  author={Victor C. Chen and Fayin Li and Shen-Shyang Ho and Harry Wechsler},
  title={Micro-Doppler Effect in Radar: Phenomenon, Model, and Simulation Study},
  journal={IEEE Transactions on Aerospace and Electronic Systems},
  volume={42},
  number={1},
  pages={2--21},
  year={2006},
  month = {Jan.},
  doi={10.1109/TAES.2006.1603402}
}

@article{molchanov2014classification,
  author={Pavlo Molchanov and Kirill Egiazarian and Jaakko Astola and Roland I. A. Harmanny and J. J. M. de Wit},
  title={Classification of Small {UAV}s and Birds by Micro-{D}oppler Signatures},
  journal={International Journal of Microwave and Wireless Technologies},
  volume={6},
  number={3--4},
  pages={435--444},
  year={2014},
  month={Mar.},
  doi={10.1017/S1759078714000282}
}

@article{fioranelli2015classification,
  author={Francesco Fioranelli and Matthew Ritchie and Hugh Griffiths and Hamid Borrion},
  title={Classification of Loaded/Unloaded Micro-Drones Using Multistatic Radar},
  journal={Electronics Letters},
  volume={51},
  number={22},
  pages={1813--1815},
  year={2015},
  month={Oct.},
  doi={10.1049/el.2015.3038}
}

@book{skolnik2008radar,
  author={Merrill I. Skolnik},
  title={Radar Handbook},
  edition={3},
  publisher={McGraw-Hill},
  year={2008}
}

@article{pu2026stf,
  author  = {Henglin Pu and Xuefeng Wang and Lu Su and Husheng Li},
  title   = {Space-Time-Frequency Synthetic Integrated Sensing and
             Communication Networks},
  journal = {IEEE Transactions on Wireless Communications},
  volume  = {25},
  pages   = {15004--15020},
  year    = {2026},
  month = {Apr.},
  doi     = {10.1109/TWC.2026.3681132}
}

@article{singh2018blade,
  author  = {A. K. Singh and Y. H. Kim},
  title   = {Accurate Measurement of Drone's Blade Length and Rotation Rate Using Pattern Analysis With {W}-Band Radar},
  journal = {Electronics Letters},
  volume  = {54},
  number  = {8},
  pages   = {523--525},
  year    = {2018},
  month = {Apr.},
  doi     = {10.1049/el.2017.4494}
}

@article{kang2021spectrum,
  author  = {K.-B. Kang and J.-H. Choi and B.-L. Cho and J.-S. Lee
             and K.-T. Kim},
  title   = {Analysis of Micro-{D}oppler Signatures of Small {UAV}s Based
             on {D}oppler Spectrum},
  journal = {IEEE Transactions on Aerospace and Electronic Systems},
  volume  = {57},
  number  = {5},
  pages   = {3252--3267},
  year    = {2021},
  month = {Apr.},
  doi     = {10.1109/TAES.2021.3074208}
}

@article{gong2022dwell,
  author  = {J. Gong and J. Yan and D. Li and D. Kong},
  title   = {Detection of Micro-Doppler Signals of Drones Using Radar
             Systems With Different Radar Dwell Times},
  journal = {Drones},
  volume  = {6},
  number  = {9},
  pages   = {262},
  year    = {2022},
  month = {Sept.},
  doi     = {10.3390/drones6090262}
}

@article{ritchie2017multistatic,
  author  = {M. Ritchie and F. Fioranelli and H. Borrion and
             H. Griffiths},
  title   = {Multistatic Micro-Doppler Radar Feature Extraction for
             Classification of Unloaded/Loaded Micro-Drones},
  journal = {IET Radar, Sonar \& Navigation},
  volume  = {11},
  number  = {1},
  pages   = {116--124},
  year    = {2017},
  month = {Apr.},
  doi     = {10.1049/iet-rsn.2016.0063}
}

@article{costa2025multipropeller,
  author  = {H. C. A. Costa and S. J. Myint and C. Andrich and
             S. W. Giehl and M. Engelhardt and C. Schneider and
             R. S. Thom{\"a}},
  title   = {Modeling Micro-Doppler Signature of Multi-Propeller
             Drones in Distributed {ISAC}},
  journal = {IEEE Journal of Selected Topics in Electromagnetics,
             Antennas and Propagation},
  volume  = {1},
  number  = {1},
  pages   = {208--222},
  year    = {2025},
  month = {Sept.},
  doi     = {10.1109/JSTEAP.2025.3604407}
}

@inproceedings{ma2024frames,
  author    = {D. Ma and J. Wei and Q. Zhang and Z. Wei and W. Jiang
               and N. Guo and S. Huang},
  title     = {Performance Evaluation of Micro-{D}oppler Based {UAV} Identification Using Different 5{G} Frame Structures},
  booktitle = {2nd International Conference on Mobile Internet,
               Cloud Computing and Information Security (MICCIS)},
  pages     = {173--179},
  year      = {2024},
  month = {Apr.},
  address = {Changsha City, China},
  doi       = {10.1109/MICCIS63508.2024.00037}
}

@article{yin2025multipath,
  author  = {M. Yin and M. Huang and R. Hu and H. Xu},
  title   = {Deep Learning-Assisted {ISAC} for {UAV} Detection in Multipath Environments With Micro-{D}oppler},
  journal = {IEEE Wireless Communications Letters},
  volume  = {14},
  number  = {9},
  pages   = {2862--2866},
  year    = {2025},
  month = {Jun.},
  doi     = {10.1109/LWC.2025.3581572}
}

@article{bai2026survey,
  author  = {S. Bai and S. Li and C. Yin and W. Qu and L.-T. Hsu
             and Y. Liu and W.-H. Chen},
  title   = {{MIMO} {OFDM}-Enabled {ISAC} for Low-Altitude Non-Cooperative
             {UAV} Surveillance: A Survey},
  journal = {arXiv preprint arXiv:2604.02680},
  year    = {2026},
  month = {Apr.},
  doi     = {10.48550/arXiv.2604.02680}
}

@article{dan2019lte,
  author  = {Yangpeng Dan and Jianxin Yi and Xianrong Wan and Yunhua Rao and Benjing Wang},
  title   = {{LTE}-Based Passive Radar for Drone Detection and Its Experimental Results},
  journal = {The Journal of Engineering},
  volume  = {2019},
  number  = {20},
  pages   = {6910--6913},
  year    = {2019},
  month = {Sept.},
  doi     = {10.1049/joe.2019.0583}
}

@article{geng2020lte,
  author  = {Zhe Geng and Renhui Xu and Hai Deng},
  title   = {{LTE}-Based Multistatic Passive Radar System for {UAV} Detection},
  journal = {IET Radar, Sonar \& Navigation},
  volume  = {14},
  number  = {7},
  pages   = {1088--1097},
  year    = {2020},
  month = {Jul.},
  doi     = {10.1049/iet-rsn.2019.0452}
}

@article{zhang2024perceptive,
  author  = {Yue Zhang and Hangguan Shan and Hongbin Chen and De Mi and Zhiguo Shi},
  title   = {Perceptive Mobile Networks for Unmanned Aerial Vehicle Surveillance:
             From the Perspective of Cooperative Sensing},
  journal = {IEEE Vehicular Technology Magazine},
  volume  = {19},
  number  = {2},
  pages   = {60--69},
  month   = jun,
  year    = {2024},
  doi     = {10.1109/MVT.2024.3373931}
}

@article{song2025overview,
  author  = {Yuxuan Song and Yong Zeng and Yuhang Yang and Zixiang Ren
             and Gaoyuan Cheng and Xiaoli Xu and Jie Xu and Shi Jin and Rui Zhang},
  title   = {An Overview of Cellular {ISAC} for Low-Altitude {UAV}:
             New Opportunities and Challenges},
  journal = {IEEE Communications Magazine},
  volume  = {63},
  number  = {12},
  pages   = {88--95},
  year    = {2025},
  month = {Jul.},
  doi     = {10.1109/MCOM.002.2400742}
}

@INPROCEEDINGS{blandino2025airborne,
  author={Blandino, Steve and Golmie, Nada and Sahoo, Anirudha and Nguyen, Thao and Ropitault, Tanguy and Griffith, David and Sonny, Amala},
  booktitle={IEEE Military Communications Conference (MILCOM)}, 
  title={Detecting Airborne Objects with {5G NR} Radars}, 
  year={2025},
  month = {Oct.},
  volume={},
  address = {Los Angeles, CA, USA},
  number={}}

@article{blandino2026gnb,
  author  = {Steve Blandino and Neeraj Varshney and Jian Wang and Jack Chuang
             and Camillo Gentile and Nada Golmie},
  title   = {Evaluation of {gNB} Monostatic Sensing for {UAV} Use Case},
  journal = {arXiv preprint arXiv:2604.02205},
  year    = {2026},
  month = {Apr.},
  doi     = {10.48550/arXiv.2604.02205}
}

@inproceedings{ji2025doppler,
  author    = {Chenqing Ji and Qionghui Liu and Jiahong Liu and Chao Yu
               and Yifei Sun and Rui Wang and Fan Liu},
  title     = {Doppler-Based Multistatic Drone Tracking via Cellular Downlink Signals},
  booktitle = {2025 IEEE Globecom Workshops (GC Wkshps)},
  year      = {2025},
  month = {Dec.},
  address = {Taipei, Taiwan},
}

@inproceedings{meng2026needle,
  author    = {Chengzhen Meng and Chenming He and Yidong Jiang and Xiaoran Fan
               and Dequan Wang and Lingyu Wang and Jianmin Ji and Yanyong Zhang},
  title     = {Needle in a Haystack: Tracking {UAVs} from Massive Noise
               in Real-World {5G-A} Base Station Data},
  booktitle = {Proceedings of the 24th Annual International Conference
               on Mobile Systems, Applications and Services},
  series    = {MobiSys '26},
  pages     = {14--27},
  year      = {2026},
  month = {Jun.},
  doi       = {10.1145/3745756.3809186}
}

@article{Luo2013,
  author  = {Ying Luo and Qun Zhang and Cheng-Wei Qiu
             and Song Li and Tat Soon Yeo},
  title   = {Three-Dimensional Micromotion Signature
             Extraction of Rotating Targets in
             {OFDM}-{LFM} {MIMO} Radar},
  journal = {Progress In Electromagnetics Research},
  volume  = {140},
  pages   = {733--759},
  year    = {2013},
  month = {Jan.},
  doi     = {10.2528/PIER13042202}
}

@article{Xu2017,
  author  = {Dan Xu and Bo Tian and Cunqian Feng
             and Sisan He and Shuang Zhao},
  title   = {Space Position Reconstruction of Ballistic
             Target Based on Adaptive Fusion},
  journal = {Journal of Beijing University of
             Aeronautics and Astronautics},
  volume  = {43},
  number  = {6},
  pages   = {1216--1223},
  year    = {2017},
  month ={Jun.},
  doi     = {10.13700/j.bh.1001-5965.2016.0463},
  note    = {In Chinese}
}

@article{Zhao2016MultiAspect,
  author  = {Shuang Zhao and Weihong Lu and
             Cunqian Feng and Sisan He and Jingqing Li},
  title   = {Feature Extraction of Precession Targets
             Based on Multi-aspect Micro-Doppler Fusion},
  journal = {Journal of Signal Processing},
  volume  = {32},
  number  = {3},
  pages   = {296--303},
  year    = {2016},
  month = {May},
  doi     = {10.16798/j.issn.1003-0530.2016.03.006},
  note    = {In Chinese}
}

@ARTICLE{8999605,
  author={Liu, Fan and Masouros, Christos and Petropulu, Athina P. and Griffiths, Hugh and Hanzo, Lajos},
  journal={IEEE Transactions on Communications}, 
  title={Joint Radar and Communication Design: Applications, State-of-the-Art, and the Road Ahead}, 
  year={2020},
  month = {Feb.},
  volume={68},
  number={6},
  pages={3834-3862},
  doi={10.1109/TCOMM.2020.2973976}}

\end{document}